\documentclass[aps,prb,twocolumn,epsfig]{revtex4}
\usepackage{epsfig}

\begin{document}

\title{Instabilities and Insulator-Metal transitions in Half-Doped Manganites
induced by Magnetic-Field and Doping}

\author{O. C\'epas,$^{a,b,*}$ H. R. Krishnamurthy,$^{a,c}$ and T. V. Ramakrishnan$^{a,c,d}$}

\affiliation{
a. Department of Physics, Indian Institute of Science, Bangalore 560012, India. \\
b. Institut Laue Langevin, BP 156, 38042 Grenoble, France.\\
c. Jawaharlal Nehru Centre for Advanced Scientific Research, Jakkur, Bangalore 560 064, India. \\
d. Department of Physics, Banaras Hindu University, Varanasi
221005, India.}

\date{\today}

\begin{abstract}
We discuss the phase diagram of the two-orbital model of half-doped
manganites by calculating self-consistently the Jahn-Teller (JT)
distortion patterns, charge, orbital and magnetic order at zero
temperature. We analyse the instabilities of these phases caused by
electron or hole doping away from half-doping, or by the application
of a magnetic-field.  For the CE insulating phase of half-doped
manganites, in the intermediate JT coupling regime, we show that there
is a competition between canting of spins (which promotes mobile
carriers) and polaronic self-trapping of carriers by JT defects. This
results in a marked particle-hole asymmetry, with canting winning only
on the electron doped side of half-doping. We also show that the CE
phase undergoes a first-order transition to a ferromagnetic metallic
phase when a magnetic-field is applied, with abrupt changes in the
lattice distortion patterns. We discuss the factors that govern the
intriguingly small scale of the transition fields. We argue that the
ferromagnetic metallic phases involved have two types of charge
carriers, localised and band-like, leading to an effective two-fluid
model.
\end{abstract}

\pacs{PACS numbers: }

\maketitle

\section{Introduction}
\label{Introduction-s}

"Half-doped" manganites, corresponding to the general formula
Re$_{1-x}$A$_{x}$MnO$_3$ with $x=1/2$ where Re is a 3+ rare-earth
metal ion and A a 2+ alkaline earth metal ion, eg.,
La$_{1/2}$Ca$_{1/2}$MnO$_3$, have been the object of extensive
experimental and theoretical studies for many
years.\cite{Salamon,LP-N} Here each Mn has an average valence of
$3.5+$ i.e., an average configuration of $d^{3.5}$, corresponding to
one Mn-$e_g$ electron for every two Mn sites hopping around amongst
the two [$(x^2-y^2)$ and $(3z^2-r^2)$] $e_g$ orbitals on each Mn. The
remaining three $t_{2g}$ electrons on each Mn are spin-aligned by
strong correlations (Hund's rules) to form "core spins" with $S=3/2$.
Similarly to the end members, i.e., LaMnO$_3$ or CaMnO$_3$, the
half-doped compounds, thanks to their commensurate filling, are
simpler in some ways than the doped manganites for general
$x$.\cite{Salamon,LP-N} Nevertheless, they exhibit a very rich variety
of properties as well.\cite{Salamon} Specifically, as the system is
cooled, there are phase transitions involving changes in magnetic,
charge and orbital order, and, in some cases, metalicity. The details
vary from material to material, depending systematically on the sizes
of the "A site" ions of the perovskite structure. Generally, the
lowest temperature phase is insulating, with simultaneous charge,
orbital and CE type antiferromagnetic order (see below), and the
charge/orbital order sets in first, at higher temperatures, compared
to the antiferromagnetic order (\textit{e.g.}, for PrCa $T_{CO/OO}
\sim 240 K$, whereas $T_N \sim 170 K$\cite{Martin}). The NdSr and PrSr
systems show ferromagnetic metallic order at intermediate
temperatures, but in the LaCa and PrCa systems, the different phases
obtained with increasing temperature continue to be
insulating. Typically, the charge order and insulating behaviour at low
temperatures persist on the "over-doped" side ($x > 1/2$), whereas the
charge order disappears rather quickly on the "under-doped" side ($x <
1/2$), and is often accompanied by metalicity (except in the PrCa
system, which stays insulating for all $x$).  This asymmetry between
"electron doping" and "hole doping" away from half-doping is a
striking feature of the majority of the half-doped manganites.

\begin{figure}[htbp]
\centerline{ \psfig{file=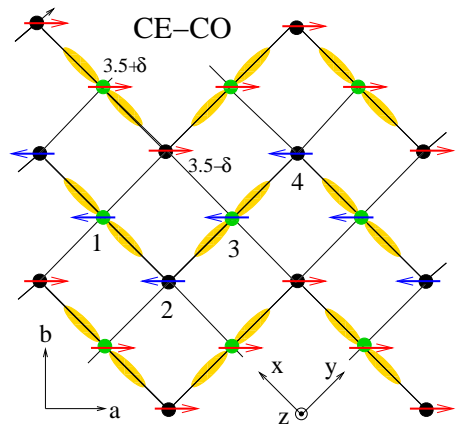,width=6.0cm,angle=-0}}
\caption{(color online). A depiction of the CE charge-ordered
antiferromagnetic phase (CE-CO). The "bridge sites" (1,3), at the
centers of the arms of the zig-zag chains which are ferromagnetically
ordered, have alternate occupancies of the $3x^2-r^2$ and $3y^2-r^2$
orbitals. At the "corner sites" (2,4), there is no
orbital order, unless JT interactions are present. $\delta$ is the
charge disproportionation between the occupancies of the corner sites
and the bridge sites.} \label{CE-phase-f}
\end{figure}

One simplifying feature of the half-doped manganites is that the
low temperature phase is generally regarded as reasonably well
characterised. Early neutron diffraction work by Wollan and
Koehler\cite{Wollan} suggested that the magnetic structure of
La$_{1/2}$Ca$_{1/2}$MnO$_3$ can be viewed as a set of
ferromagnetic zig-zag chains antiferromagnetically ordered
relative to each other, with an 8-sublattice, ($2\sqrt{2} \times
2\sqrt{2}$) unit cell, and is referred to as the CE magnetic order
(Fig. \ref{CE-phase-f}). The structure was qualitatively explained
soon thereafter by Goodenough,\cite{Goodenough} who proposed
additionally that the phase also has a 2-sublattice, $\sqrt{2}
\times \sqrt{2}$ charge order with alternating Mn$^{3+}$ and
Mn$^{4+}$ sites, and a 4-sublattice, $2\sqrt{2} \times \sqrt{2}$
striped orbital order, as indicated in Fig. \ref{CE-phase-f}.
Since then, CE order has been found in several other half-doped
systems, such as
Nd$_{1/2}$Sr$_{1/2}$MnO$_{3}$\cite{Kawano0,Kawano} or
Nd$_{1/2}$Ca$_{1/2}$MnO$_3$\cite{Millange}, though some, such as
Pr$_{1/2}$Sr$_{1/2}$MnO$_3$,\cite{Kawano0,Kawano} show A-type
antiferromagnetism, corresponding to $[0,0,\pi]$ spin order, i.e,
ferromagnetic planes of spins which are antiferromagnetically
aligned in the z-direction).

The presence of charge and orbital order is, however, harder to
establish directly experimentally because of the lack of experimental
probes that couple directly to them. Indeed, the perfect
Mn$^{3+}$/Mn$^{4+}$ charge ordering proposed by
Goodenough.\cite{Goodenough} is currently regarded as
controversial\cite{Garcia,Daoud-Aladine,Coey,Efremov}. X-ray
diffraction data do suggest the presence of large Jahn-Teller (JT)
distortions of the oxygen octahedra surrounding the Mn
sites\cite{Radaelli,Kawano} with two inequivalent Mn sites, of
effective valence $3.5+\delta$ and $3.5-\delta$, but $\delta$ is not
really known, and is unlikely to be close to 0.5. In
Pr$_{0.6}$Ca$_{0.4}$MnO$_3$ (which is slightly under-doped though),
the charge or valence contrast seems further reduced, and it has been
suggested on the basis of neutron diffraction data that the electron
is rather shared by two Mn sites paired in dimer-like structures,
referred to as "Zener Polarons", with ($\delta \ll
0.5$).\cite{Daoud-Aladine} It is not clear, however, whether this is
due to the presence of additional electrons\cite{noteKhomskii} or a
general feature of many half-doped manganites.\cite{Daoud-Aladine2} A
recent work on Pr$_{0.5}$Ca$_{0.5}$MnO$_3$ claims indeed to confirm
the picture of the original CE state at precisely
half-filling.\cite{Goff} An alternate, closely related picture of the
half-doped system is that of a bond-charge-density-wave, with no
charge contrast of the Mn ions, substantial hole occupancy on the
oxygen ions on the chains, and alternating "O$^{2-}$/O$^{-}$"
order.\cite{Zheng} There have been X-ray studies on orbital order and
correlations as well as charge and magnetic order using a variety of
methods such as soft x-ray resonant diffraction,\cite{thomas} coherent
x-ray scattering,\cite{nelson} which explore the spatial extent of
orbital order, in particular. The resonant scattering experiments in
Pr$_{0.6}$Ca$_{0.4}$MnO$_3$ conclude that the charge
disproportionation is less than complete, that there is orbital
mixing, and that therefore the simple Goodenough model is not
right. On the other hand, when holes are in excess ($x > 0.5$), it has
been suggested that charge order persists but becomes
incommensurate.\cite{incom-CO,Brey}

A closely related family of manganites carefully studied recently is
A$_{0.5}$A'$_{0.5}$MnO$_3$ where A is a rare earth (Y, Tb, Sm, Nd, Pr,
La in order of ion size) and A' is Ba.\cite{F_M} The phases have been
studied as a function of A-A' site order/disorder. When there is
order, the low temperature phase is charge-ordered (CO) and orbitally
ordered (OO) for ion size from Y to Nd. Beyond Nd, up to La, the phase
is a ferromagnetic metal. However, when A-A' sites are disordered,
there is no CO phase at all, but only a spin glass (SG) phase, from Y
to Sm, after which the ground state is ferromagnetic (FM). This means
that the CO/FM and CO/SG competition depends on ion size as well on A
site ordering in the perovskite ABO$_3$ structure.

Another interesting and intriguing property of the CE charge-ordered
(CE-CO) phase is the magnetic-field-induced insulator-metal transition
first discovered in
(Nd,Sm)$_{1/2}$Sr$_{1/2}$MnO$_3$,\cite{Tokura,TokuraTomioka} and later
shown to be ubiquitous\cite{Salamon}. Though insulating at zero field,
these materials become metallic by the application of magnetic-fields
of the order of 10 - 40 Tesla via sharp, first-order, resistive
transitions.\cite{Tokura} The magnetic-field energies involved are
much smaller than the thermal energies (of order 200 K) needed to
destroy the charge order, and orders of magnitude smaller than the
charge gap of 0.2 - 0.3 eV, as observed as a function of field by STM
in Nd$_{0.5}$Sr$_{0.5}$MnO$_3$.\cite{amlan} This can be viewed as a
different manifestation of the colossal magneto-resistance seen at the
metal-insulator transition of doped manganites for $x \sim
0.25$,\cite{Salamon} and the microscopic understanding of the above
features poses similar difficult theoretical
challenges.\cite{Salamon,LP-N}

\textit{Theory}. A variety of models and mechanisms have been examined
in the context of half-doped manganites as well.\cite{Salamon} The
simplest model has mobile electrons moving amongst
\textit{non-degenerate} orbitals, coupled to the Mn $t_{2g}$ core
spins by a large Hund's rule (double) exchange coupling $J_H$. The
latter promotes ferromagnetism, but competes with a direct
antiferromagnetic coupling $J_{AF}$ between the core spins. Even in
this simple model, the ferromagnetic or CE types of order are
stabilised depending upon the strength of
$J_{AF}$.\cite{Pandit,Fratini} Van den Brink \textit{et al.}
considered a more realistic model with the two types of $e_g$ orbitals
of Mn, and found that the CE phase is orbitally-ordered: the "bridge
sites" of the zig-zag chains have alternating preferred occupancy of
$(3x^2-r^2)$ and $(3y^2-r^2)$ orbitals\cite{Khomskii} as indicated in
Fig. \ref{CE-phase-f}.  They also showed that a charge contrast
$\delta$ (not bigger than 0.2) can be generated by including on-site
Coulomb interaction. This is because the "corner sites" turn out to
have equal occupancy of $(x^2-y^2)$ and $(3z^2-r^2)$ orbitals, and
that costs Coulomb energy. To reduce this, the system adopts a
preferred occupancy of the bridge sites which are orbitally ordered.
In later work, nearest neighbour Coulomb interactions were also
included.\cite{Jackeli,Shu} However, the charge order due to
long-range Coulomb interactions is generally of the Wigner-type with
wave vector $Q=(\pi,\pi,\pi)$, contrary to the $(\pi,\pi,0)$ order,
with charge stacking along the z-direction, suggested by
experiment. To stabilise the $(\pi,\pi,0)$ order in a wider regime of
parameters, JT interactions between the Mn ions and their surrounding
oxygen octahedra, which are supposed to be quite large,\cite{Radaelli}
have to be included.\cite{Dagotto} The consequent JT distortions
further lower the energy of the CE phase because of the already
present $(3x^2-r^2)/(3y^2-r^2)$ orbital order.  Classical Monte Carlo
simulations including static JT distortions on small clusters as well
as self-consistent mean field treatments of models including JT and
Coulomb interactions\cite{Dagotto} suggest that the CE charge stacked
state has the lowest energy in an intermediate range of $J_{AF}$,
unless the nearest neighbour coulomb interaction $V$ becomes much too
large.

However, to our knowledge very few of these studies have addressed the
other issues, such as the magnetic-field-induced insulator metal
transition, and the electron-hole asymmetry. The first issue was
tackled in refs. [\onlinecite{Pandit,Fratini,Dagotto2}] by assuming
model parameters very close to the phase boundary between the
ferromagnetic and CE states. The resulting small energy difference between
the two phases can then be overcome by an arbitrary small
magnetic-field. But it is hard to justify why the system parameters
should be so finely tuned for so many systems. As regards the second
issue, band structure arguments,\cite{Khomskii} and treatments
including JT distortions on small clusters\cite{Dagotto} necessarily
lead to metallic phases upon addition of electrons or holes, in
contrast to experiments.

Recently, a theory for doped manganites has been proposed\cite{Ram}
where it is argued that due to strong JT interactions the $e_g$
electrons dynamically reorganise themselves into two types. The
majority of the electrons (labelled $\ell$) become localised polarons,
trapped by large local JT distortions; and a minority of them
(labelled $b$) can remain mobile and non-polaronic.  Still virtual
adiabatic transitions to empty neighbouring sites induces a
ferromagnetic exchange referred as \textit{virtual} double
exchange.\cite{Ram} The resulting Falicov-Kimball like, $\ell-b$ model
Hamiltonian treated in a simple dynamical mean-field treatment in the
framework of an "orbital liquid" description, gave a good account of
the properties of doped manganites.\cite{Ram}

In this paper, we propose an extension of the above theory to the half
doped case, which has to include the possibilities for orbital,
charge, and antiferromagnetic order. We obtain pointers to this by
studying the properties of electronic excitations {\it coupled with JT
defects} in the lattice distortion pattern. We find that such a study
suggests the incipient instabilities of the CE phase indicative of the
doping and magnetic-field induced phase transitions seen
experimentally, as well as the presence of localised and mobile
carriers. The localised states on the defects which we obtain are
different from the microferrons suggested at small $x$ around a
dopant,\cite{LP-N} as they are self-generated and could exist even in
the absence of chemical disorder. In principle, the JT defects we are
considering could be mobile on a longer time-scale, although disorder
may indeed pin them down.

More specifically, in this paper, we first determine the zero
temperature phase diagram of the 3d two-orbital model of half-doped
manganites for periodic phases in the {\it thermodynamic limit},
including JT distortions, but ignoring Coulomb interactions. We do
this by minimising the energy assuming a periodic unit-cell of eight
sites,\cite{NoteCE} inside which static JT distortions and core spin
directions are allowed to be arbitrary. This allows us to determine
them self-consistently without using finite-size clusters, thereby
extending and reinforcing earlier work.\cite{Dagotto,brey-pd} In
particular we obtain analytic results for the phase boundaries at
strong JT coupling.

Next, we show that the periodic ferromagnetic phase obtained at small
$J_{AF}$ by the method discussed above can become unstable with
respect to a phase with two types ($\ell-b$) of electrons when the JT
coupling is lowered. We show indeed that it becomes energetically
favourable to create single site JT defects, i.e.  release the
distortion on a finite number of sites that were previously distorted
and promote previously trapped electrons onto a mobile band, thus
suggesting a metallic phase.  The exact nature of the phase can not be
figured out by such an instability analysis. Nonetheless, it suggests
an effective ($\ell-b$) Hamiltonian with orbital degrees of freedom
explicitly included.

The observed phases at half-doping, such as the CE phase, are
antiferromagnetic, corresponding to appropriately larger values of
$J_{AF}$. But they show transitions to ferromagnetic metallic phases
in an external magnetic-field or when electrons are added. To
understand such transitions, in addition to considering changes in the
JT distortions, canting of spins is important.

Canted phases are expected to appear not only in a magnetic-field, but
also upon doping with carriers (and irrespective of their nature),
following the original argument by de Gennes.\cite{deGennes} Here we
show, however, that canted metallic phases appear only when
\textit{electrons} (and not holes) are added, because of the
underlying asymmetry of the density of states at half-doping. When we
allow for JT distortions, we find a competition with a disordered
phase where the added electrons are trapped by JT distortions, the
latter phase winning only at small electron concentration.  On the
hole-doped side, added holes are simply trapped by the lattice
distortions and the system remains insulating. Thus our work provides
an explanation for the particle-hole asymmetry near $x \sim 1/2$, at
intermediate JT couplings which we argue are relevant for the majority
of the manganites.

Similarly, we find that there is a strong interplay between turning on
a magnetic-field at half doping and the JT distortion pattern. This is
consistent with x-ray measurements in La$_{1/2}$Ca$_{1/2}$MnO$_3$ in a
field.\cite{Tyson,Nojiriprivate} Starting from the distorted CE phase,
we find in addition an instability of the high-field ferromagnetic
phase to the formation of JT defects.  This suggests that the
high-field phase seen in experiments may need a two-fluid description.

Interestingly, a very similar two-carrier hypothesis was proposed
based on phenomenological grounds to understand the resistivity of
La$_{1-x}$Ca$_{x}$MnO$_3$ ($x \sim 1/2$).\cite{Roy} More recently, a
particle-hole asymmetric Ginzburg-Landau theory was proposed to
explain\cite{milward} the incommensurate charge order\cite{incom-CO}
seen for $x > 0.5$. We believe that our theory provides the
microscopic basis for these facts both. 

The rest of this paper is organised as follows. In section \ref{pd-s},
we discuss the phase diagram of the half-doped manganites restricted
to periodic ground states with the most general 8-sublattice
structure. We give in particular an analytic strong-coupling
description (\ref{wannier-ss}). In section \ref{Instabilities-s}, we
study the instabilities of some of these phases : instability of the
strong JT coupling ferromagnetic phase, which defines a new phase
(\ref{InstabilityFerro-ss}); instability upon doping to the canted
phases (\ref{doping-ss-canted}) or to self-trapping of added carriers
(\ref{doping-ss-trapped}), and the competition between the two
(\ref{competionbetweenthetwo}). The effect of the magnetic-field on
the CE phase is studied in section \ref{mag-f-ss}, where we also
discuss the nature of the high-field ferromagnetic phase.  In section
\ref{Conclusion-s}, we summarise and discuss the possibilities for
testing these ideas experimentally. A short account of some of these
results has been presented elsewhere.\cite{short}

\section{Optimised Periodic Phases and Phase Diagram for Half Doped Manganites}
\label{pd-s}

\subsection{Model Hamiltonian}
\label{mh-ss}

Our discussions are based on the following Hamiltonian for the
manganites

\begin{eqnarray}
{\cal H}[\{ \textbf{\mbox{S}}_{ia} , Q_{i a},\Theta_{i a} \} ] = -
\sum_{ij \alpha \beta ab} \tilde{t}_{ab ij}^{\alpha \beta}
(\textbf{\mbox{S}}_{ia},\textbf{\mbox{S}}_{jb} ) c^{\dagger}_{i a
\alpha } c_{j b \beta} \nonumber \\+ \sum_{<ijab>} J_{AF}
\textbf{\mbox{S}}_{ia} . \textbf{\mbox{S}}_{jb} - \mathrm{g} \mu_B
\sum_{ia} \textbf{\mbox{H}}
\cdot \textbf{\mbox{S}}_{ia}  \nonumber \\
+ \frac{1}{2} K \sum_{i a} Q_{i a}^2 - g \sum_{i a \alpha \beta
} Q_{i a} \tau_{\alpha \beta}(\Theta_{i a})  c_{i a \alpha
}^{\dagger} c_{i a \beta}\label{Hamiltonian}
\end{eqnarray}
where $c^{\dagger}_{i a \alpha }$ creates an $e_g$ electron (in a
low-energy-projected Wannier orbital with $e_g$
symmetry\cite{low-en-p-H}), on the sublattice site $a$ (Mn site) of
the unit-cell $i$ in the 3d cubic lattice, and in the orbital state
$\alpha=1,2$, with $1 \equiv d_{x^2-y^2}, 2 \equiv~d_{3z^2-r^2}$. The
original lattice is decomposed into eight sublattices, labelled with
$a$.\cite{NoteCE} There are $N$ sites and $cN = (1-x)N$ electrons
(when $x=1/2$ the number of electrons is denoted $N_0 \equiv N/2$).
The first term is the kinetic energy of the electrons. The hopping
parameters are taken to be of the usual Anderson-Hasegawa
form\cite{AndersonHasegawa} which takes care of the Hund's coupling,
$J_H \sum_i \textbf{\mbox{S}}_{ia} \cdot \textbf{\mbox{s}}_{ia} $ in
the limit of large $J_H/t$, with $\textbf{\mbox{S}}_{ia}$, the $S=3/2$
core spin formed from the Mn $t_{2g}$ electrons being approximated as
a classical spin. As a consequence only the electrons with spin
projections parallel to the core spins are present, and their hopping
amplitudes are functions of the polar angles of the core spins given
by:\cite{AndersonHasegawa,LP-comm}
\begin{eqnarray}
&\tilde{t}_{ab ij} ^{\alpha \beta}&(\textbf{\mbox{S}}_{ia},
\textbf{\mbox{S}}_{jb}) = t_{ab ij}^{ \alpha  \beta}
\times \nonumber \\
&\times& \left( \cos
\frac{\theta_{ia}}{2}  \cos
\frac{\theta_{jb}}{2} + \sin
\frac{\theta_{ia}}{2}  \sin
\frac{\theta_{jb}}{2} e^{i(\phi_{ia}-\phi_{jb})} \right)
\end{eqnarray}
Here $t_{ab ij}^{ \alpha \beta}$ is the usual, anisotropic and
symmetry determined, hopping amplitude\cite{Dagotto} between the
$e_g$ orbitals $\alpha$ and $\beta$ at the two nearest neighbour
sites $(i,a)$ and $(j,b)$ respectively, arising from their
hybridisation with the $O-p_\sigma$ orbitals (with $4t/3$ being
the hopping between $(3z^2-r^2)$ orbitals in the
$z$-direction)\cite{hopping}. The second term $J_{AF}$ is the
antiferromagnetic coupling of the $t_{2g}$ core spins that comes
from standard superexchange processes.\cite{Millis-SE} It can be
roughly estimated from the N\'eel temperature of a system with
only $t_{2g}$ core spins, such as CaMnO$_3$, although the
structure of the half-doped system is not exactly the same. The
third term is the Zeeman energy where $\textbf{\mbox{H}}$ is the
external magnetic-field. The last two terms include the
Jahn-Teller (JT) phonons and their coupling to the $e_g$
electrons.  We neglect the $P_{i a}^2/2M_{ia}$ terms since $\hbar
\omega_0 \ll t$ (where $\omega_0$ is the typical phonon
frequency), but {\it include their effects heuristically when we
argue that JT defects lead to polaron formation}. $Q_{i a }$ and
$\Theta_{i a}$ are, respectively, the amplitude (measured in units
of the typical JT distortions in these materials) and the angle of
the JT distortion at the site $(i,a)$. The coupling matrix is
given by:

\begin{equation}
\tau(\Theta) = \left( \begin{tabular}{cc}
$\cos \Theta$ & $\sin \Theta$  \\
$\sin \Theta$ & $-\cos \Theta$
\end{tabular} \right)
\end{equation}
$K$ is the lattice stiffness of a simplified non-cooperative model
where distortions on neighbouring sites are not coupled. More
detailed and realistic models would include cooperative JT
couplings and coupling to breathing modes such as in the lattice
model of Ref. [\onlinecite{Milliscooperative}].

We have neglected the on-site Coulomb interaction, $U \sum_i n_{i
a \alpha } n_{i a \bar{\alpha}}$ between different orbital states.
Although it is an important interaction in the problem, we can not
treat it using the methods used in this paper except in a
mean-field approximation. However, it is effectively taken into
account when orbital order occurs. We comment on the effects of
its inclusion at appropriate places in the paper. When a local JT
distortion occurs on a site, the degeneracy of the $e_g$ orbitals
is lifted. If only one electron is present, there is a gain by
occupying the lowest energy level. If a second electron is added,
however, it has to occupy the higher energy level because of the
strong Hund's coupling.  There is a compensation and the energy
gain vanishes. JT distortions therefore suppress double occupancy
of sites, mimicking the effect of $U$.  When $g \ll t$, the
distortions are small or zero and it is important to explicitly
include $U$, which does play a role. For instance, it induces a
charge-ordering in the CE phase,\cite{Khomskii} just as a finite
$g/t$ does.\cite{Dagotto} When $g \gg t$ (see section
\ref{wannier-ss}), JT distorted phases appear naturally, and the
inclusion of $U$ is not crucial. Similarly, the inclusion of the
term $U \sum_i n_{i \uparrow} n_{i \downarrow}$ is unimportant
(completely irrelevant when $J_H \rightarrow \infty$) because the
large Hund's coupling prevents double occupancy of this type. We
have also neglected the longer range coulomb interactions as they
are expected to be weak because of the large dielectric constant
of the manganites, and we do not consider issues (such as
macroscopic phase separation) which are sensitive to their
presence.

Note also that regarding the direct coupling of the $t_{2g}$
spins, we restrict ourselves to a pure Heisenberg superexchange
coupling. This may not be absolutely accurate for $S=3/2$ spins.
Further couplings, such as single-ion anisotropies, are certainly
present in the real materials, but are not of crucial importance
for the issues we focus on in this paper. We therefore restrict
ourselves to the Hamiltonian of eq. (\ref{Hamiltonian}).

\subsection{Method}
\label{method-ss}

The Hamiltonian of eq. (\ref{Hamiltonian}) represents mobile
electrons coupled to local classical degrees of freedom that act
like {\it annealed disorder}. The probability weight of a
particular configuration of the classical variables is given by
$\exp[-F_{el}/(k_B T)]$ where $F_{el}$ is the electronic free
energy in the presence of that configuration; their distribution
thus has to be determined self-consistently. At zero temperature,
it is reasonable to assume that the classical degrees of freedom
are frozen and have well-defined values. On a finite lattice these
can, in principle, be determined as follows. One can diagonalise
the Hamiltonian exactly for a given (arbitrary) configuration of
lattice distortions $(Q_i,\Theta_i)$ and polar angles of the spins
$\theta_i$ (for simplicity we are ignoring the azimuthal angles of
the spins $\phi_i$). For each configuration, one can thus find the
electronic energy levels, fill up the states up to the Fermi
energy, and calculate the total energy. To obtain the ground state
of the system, one then needs to minimise this energy with respect
to all the possible configurations of classical variables. Such a
procedure can be implemented, for example, using a Monte-Carlo
technique\cite{Furukawa} for a finite lattice, but becomes a more
and more difficult task as the number of lattice sites, and hence
the number of variables, increases.

Since our aim is to explore the physics of the
experimentally observed CE state, we adopt a
simpler approach. We assume a 8-sublattice periodic structure that
is compatible with the periodicity of the CE state, which permits
us to tackle the problem in a lower dimensional space of the
classical variables. [Needless to say, this rules out the
possibility of incommensurate (with respect to the assumed eight
sublattice structure) or inhomogeneous phases.] We have
implemented a simulated annealing routine to minimise the total
energy with respect to (essentially all possible) distortions and
spin angles on the 8-sublattices. Thus our approach differs from
and is complementary to earlier numerical approaches which have
considered small clusters and done a full classical Monte-Carlo
simulation for the spin and lattice variables.\cite{Dagotto} For
us, the only limitation is the number of sublattices, which we fix
to be eight; the system size is vastly larger (we are treating the
real 3d case), and practically in the thermodynamic limit. [The
system size, i.e, the total number of sites, is 8 times the number
of periodic repetitions of blocks of 8 sites (the spin and
distortions being the same in all the blocks), and we have done
calculations using up to 6912 blocks.] The computational effort of
such an approach compared with that of Ref. [\onlinecite{Dagotto}]
is, on one hand, much reduced because we do not have to
equilibrate a large number of variables. On the other hand, the
calculation of the energy of each configuration takes more time
because we sum over a large number of k-points (equal to the
number of blocks) in the Brillouin zone corresponding to the
periodic structure. We have carefully studied the finite-size
effects on these Brillouin zone sums. The error on the total
energy due to the truncation of the sums is of the order of
$10^{-2}t$ where $t$ is the typical energy scale of the problem.
The thermodynamic limit is therefore reached within this accuracy,
i.e., for all the phases compatible with the sublattice structure
we expect that our results are within 1\% of the thermodynamic
limit results.

\subsection{Results}
\label{results-ss}

\begin{figure}[htbp]
\centerline{ \psfig{file=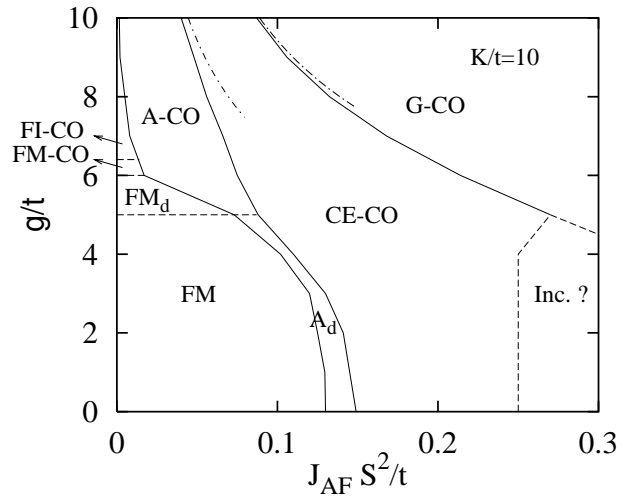,width=7cm,angle=-90}}
\caption{Phase diagram of the 3D two-orbital model ($T=0$, $x=0.5$,
$K/t=10$). FM (resp. FM$_d$): ferromagnetic metallic phase with no
distortions (resp. small uniform distortions). FI-CO (resp. FM-CO):
charge-ordered ferromagnetic insulating (resp.  metallic) phase with
distortions that favour occupancy of the $x^2-y^2$ orbitals
(Fig. \ref{pd1-f}). A$_d$: ferromagnetic planes AF aligned with
uniform distortions. A-CO: A with charge order.  CE-CO: Ferromagnetic
zig-zag chains AF ordered, charge and orbital ordered
($3x^2-r^2/3y^2-r^2$) [Fig.  \ref{CE-phase-f}]. G-CO: N\'eel AF phase
with charge-order. Inc.: Possible incommensurate states that
interpolate between CE and G. Dotted dashed lines come from analytical
expressions derived in the strong-coupling limit (section
\ref{wannier-ss}) . Solid (dashed) lines show first-order
(second-order) phase transitions.} \label{pd-f}
\end{figure}

\begin{figure}[htbp]
\centerline{ \psfig{file=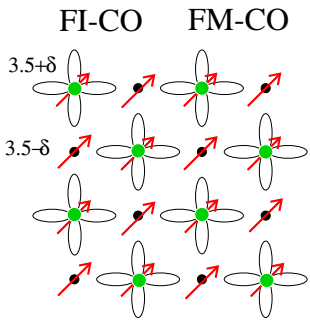,width=7.0cm,angle=-0} }
\caption{(color online). A depiction of the ferromagnetic insulating (resp.
metallic) charge-ordered phase (FI-CO [resp. FM-CO]) stable at
strong JT coupling and small antiferromagnetic coupling (see fig.
\ref{pd-f}). Alternate sites have charge disproportionation
$\delta$ (given in Fig. \ref{chargedisproportionation-f} as a
function of $g/t$). The sites with higher occupancies also have
strong JT distortions (see fig. \ref{distortions-f}), of such
orientation as to promote the occupancy only of $(x^2-y^2)$ on
these sites, leading to orbital order as well. Note that the
lattice is rotated by 45 degrees with respect to fig.
\ref{CE-phase-f}.} \label{pd1-f}
\end{figure}
Figure \ref{pd-f} shows the phase diagram as a function of the JT
coupling, $g/t$, and the antiferromagnetic coupling,
$J_{AF}S^2/t$, at zero temperature. [We choose units such that the
JT distortions are dimensionless, whence K and g both have
dimensions of energy, which we specify in units of t. We use a
fixed $K/t = 10$. To compare with earlier work, the JT energy is
then $E_{JT}/t = (g/t)^2/(2K/t) = (g/t)^2/20$.] We basically find
the same phases that were found before either by comparing the
energies of selected phases at
$g=0$,\cite{Khomskii,Jackeli,notedifference} or by Monte-Carlo
simulations at finite $g$;\cite{Dagotto} except that now we have
provided confirmation that they are indeed the optimal
8-sublattice structures in the thermodynamic limit.

We now describe the different phases shown in Fig. \ref{pd-f},
including the amplitudes of the JT distortions in them and the
corresponding electronic properties.

\begin{figure}[htbp]
\centerline{ \psfig{file=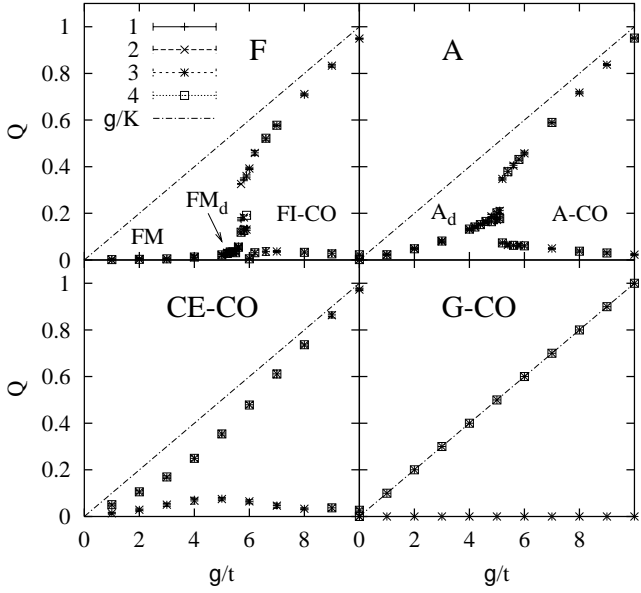,width=8.5cm,angle=-90}}
\caption{Amplitudes of distortions of the four inequivalent sites of
the unit-cell as function of the coupling parameter $g/t$. The
different panels represent the various phases found previously:
F,A,CE,G. The last three phases do not exist for all values of $g/t$;
the curves are then obtained by fixing the magnetic structure and
optimising with respect to the distortions. For the G-CO phase, the
distortions are exactly given by $g/K$ since the electrons are
completely localised.}
\label{distortions-f}
\end{figure}

\begin{figure}[htbp]
\centerline{ \psfig{file=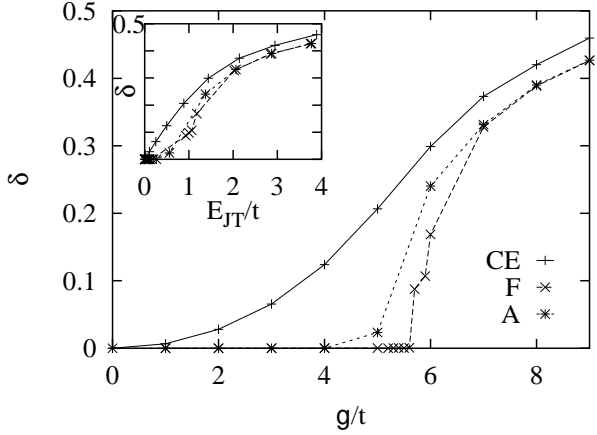,width=6cm,angle=-90}}
\caption{Charge disproportionation defined by the valence of the two
inequivalent Mn ions, Mn$^{3.5+\delta}$ and Mn$^{3.5-\delta}$ (see
Figs.  \ref{CE-phase-f} and \ref{pd1-f}), as a function of $g/t$ (or
$E_{JT}/t$ [inset] defined by $2E_{JT}=gQ_{max}$, where $Q_{max}$ is
the distortion of the site with the largest distortion) for the CE, F
and A type-phases.}
\label{chargedisproportionation-f}
\end{figure}

\begin{figure}[htbp]
\centerline{ \psfig{file=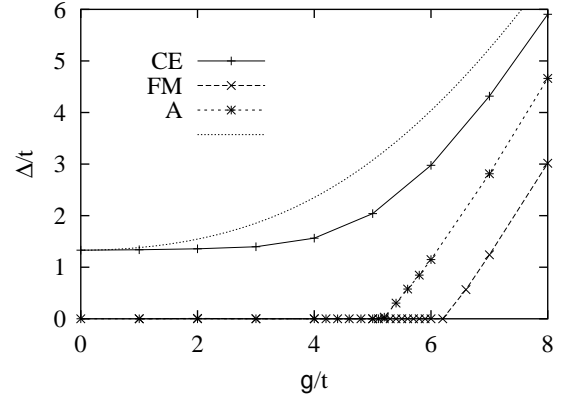,width=5.5cm,angle=-90}}
\caption{Charge gap vs. $g/t$ for the CE, F and A type-phases. The
CE phase is always insulating, while the A and F
phases are insulating beyond $g_{cA}/t=5.1$, and
$g_{cF}/t \sim 6.3$. The dotted line is the gap obtained
with non-optimised distortions (eq. (\ref{gap}) in \ref{doping-ss}).} \label{chargegap-f}
\end{figure}

For small values of $J_{AF}$ an undistorted metallic phase with
3-d ferromagnetic order (FM) is stable up to a critical value of
$g/t \sim 5$. Above this threshold, there is a ferromagnetic phase
with very small uniform distortions (Fig. \ref{distortions-f}),
noted FM$_d$. There is also a narrow region ($5.6 < g/t < 5.9$)
where the solution displays many inequivalent sites. As discussed
later, we believe that this is consistent with the instability
that we find in section \ref{Instabilities-s}. For $g \gtrsim
5.9$, the stable phase is the FM-CO followed by the FI-CO.  These
phases have the structure depicted in Fig. \ref{pd1-f}; i.e., a
layered structure with large JT distortions on two sites out of
four in a checker-board pattern in each layer, and oriented in
such a way as to favour the $(x^2-y^2)$ orbital on the strongly
distorted sites (Fig.  \ref{pd1-f}). There is a charge
disproportionation $\delta$ that is given in Fig.
\ref{chargedisproportionation-f}. The structure is metallic
(FM-CO) up to $g/t \sim 6.3$, and is insulating (FI-CO) for larger
$g/t$, as is clear from Fig. \ref{chargegap-f}. This structure has
been found before,\cite{Dagotto} and is known to compete with a
similar structure which prefers a $(3x^2-r^2)/(3y^2-r^2)$ orbital
order, when strong anharmonic and cooperative JT couplings are
taken into account.\cite{Dagotto2}

The A phases, which are more stable at larger $J_{AF}$ (the larger
the $g/t$, the smaller the coupling $J_{AF}$ required for the
transition) are similar to the ferromagnetic phases we have just
described except that successive layers are now
antiferromagnetically ordered. For $g/t$ up to $\sim 5.1$, the
phase noted A$_d$ is uniformly distorted with a distortion
amplitude given in Fig. \ref{distortions-f}. It is metallic in
this regime (see the charge gap in Fig. \ref{chargegap-f}). For
larger values of $g/t$, the A phase become charge-ordered and
insulating, as in case of the FI-CO phase (Fig. \ref{pd1-f}).

The CE-CO phase, the CE phase with charge stacked order and
orbital order (Fig. \ref{CE-phase-f}), is the stablest over a wide
range of parameters for intermediate $J_{AF}$, as is clear from
(Fig. \ref{pd-f}). As pointed out in Refs. [\onlinecite{Khomskii,
Dagotto}], there is orbital ordering even at $g=0$, but no charge
ordering; the "bridge sites" having an average occupancy of 0.5,
but only of $(3x^2-r^2)$ and $(3y^2-r^2)$ orbitals alternately
(Fig. \ref{CE-phase-f}). The "corner sites" on the other hand,
have equal occupancy (0.25 each) of $(x^2-y^2)$ and $(3z^2-r^2)$
orbitals. The sites are undistorted at $g=0$, but get distorted as
soon as $g>0$ (Fig.  \ref{distortions-f}, bottom-left panel). The
distortions on the bridge sites are the largest, and are oriented
in such a way as to further stabilise the alternating occupancy of
the $(3x^2-r^2)$ and $(3y^2-r^2)$ orbitals that already exists at
$g=0$, since distortions that precisely favour this alternation
lower the energy of the system. In addition, small distortions get
generated also on the corner sites that favour the $(x^2-y^2)$
orbital (Fig. \ref{distortions-f}). As a further consequence, a
charge disproportionation $\delta$ between the bridge and corner
sites develops, favouring a higher occupancy of the former. The
variation of $\delta$ with $g/t$ is shown in Fig.
\ref{chargedisproportionation-f}. The system is an insulator
whatever the charge disproportionation, as shown by the finite
charge gap in Fig. \ref{chargegap-f}.

For strong JT coupling ($g/t \gg 1$), the CE-CO phase is
degenerate energetically with the C-CO phase which consists of
straight ferromagnetic chains antiferromagnetically ordered with
respect to each other. The charge order is accompanied by orbital
order of the $(3z^2-r^2)$ type if the chains are oriented along
the $z$-direction. This degeneracy is discussed in explicit detail
in section \ref{wannier-ss}.  In this limit, it is easy to show
(see section \ref{wannier-ss}) that the G phase (completely 3d-AF
phase with localised electrons) is always the stablest for large
values of $J_{AF}$. For smaller values of $g/t$ (and large
$J_{AF}$) we find solutions with many inequivalent (canted) sites,
suggesting that the transition from the CE phase to the G phase in
this regime might proceed via intermediate states that are
incommensurate relative to the periodicity of the unit-cell we
have considered (Fig. \ref{pd-f}).

We consider next the strong JT coupling limit, $g/t \gg 1$, whence
we can calculate the energetics of the phases and the phase
boundaries discussed above analytically.

\subsection{Localised description for $g/t \gg 1$}
\label{wannier-ss}

At large $g/t$, it is energetically favourable for all the
electrons in the system to be self-trapped by local lattice
distortions since the JT energy gain is large. We hence
start with Wannier-type wave-functions with electrons fully
localised on strongly distorted sites. The local energy per
electron is the sum of the elastic energy $\frac{1}{2} KQ^2$ and
the electronic energy gain $-gQ$ which is minimal at $Q=g/K$ with
a net energy gain of $E_{JT} = {g^2}/(2K)$. In the limit of large
$g/t$, there is a large degeneracy because electrons can be
trapped on any site and in any orbital state, as long as the
orbital state correlates with the orientation $\Theta_i$ of the JT
distortion as
\begin{equation}
|\Psi( \Theta_i) \rangle =\cos \frac{\Theta_i}{2} |d_{x^2-y^2} \rangle
+ \sin \frac{\Theta_i}{2} |d_{3z^2-r^2} \rangle
\label{orbitaldefinition}
\end{equation}
(for instance, $\Theta_i=-\pi/3$ for $3x^2-r^2$, and
$\Theta_i=\pi/3$ for $3y^2-r^2$).

This degeneracy is lifted at second-order in perturbation theory
in the kinetic energy of the electrons. Consider an electron
localised on a site with an empty neighbouring site and with the
corresponding core spins aligned. Then, in the adiabatic limit ($t
\gg \hbar \omega_0$ where $\omega_0$ is the frequency of the JT
phonons) appropriate here, it can hop virtually 
onto any of the two orbital states of that site and
back, without giving the lattice distortions time to relax (the
relevant energy denominator being $2E_{JT}$), and hence lower its
energy. This energy lowering is less if the core spins are
misaligned whence the hopping amplitude is reduced (and even fully
suppressed in case of anti-alignment). It is also less if the
neighbouring site is occupied, whence the energy denominator is
larger, equal to $4E_{JT}$ ($4E_{JT}+ U$ in the presence of $U$,
so that the process gets suppressed altogether for large $U$).
Such a process hence gives rise to a effective double exchange
term in the Hamiltonian as pointed out in ref. [\onlinecite{Ram}]
and labelled virtual double exchange. The dominant term in the
effective Hamiltonian is then\cite{noteneglectedterms}:

\begin{eqnarray}
&&  \tilde{\cal H} = - E_{JT} \sum_i n_i + \sum_{<ij>}  J_{AF}
\textbf{\mbox{S}}_i . \textbf{\mbox{S}}_j - \mathrm{g} \mu_B
\sum_{i}
\textbf{\mbox{H}} \cdot \textbf{\mbox{S}}_i \nonumber \\
&& - {\frac {J}{2}} \sum_{<i,j>} (\textbf{\mbox{S}}_i \cdot
\textbf{\mbox{S}}_j+S^2) \left[ n_i(1-n_j) C_{i,j}^2
 + ( i \leftrightarrow j)\right]  \label{effectiveE}
\end{eqnarray}
Here $E_{JT}={g^2}/(2K)$, $n_i$ is the electronic occupancy on
site $i$, $J= {\tilde{t}^2}/(2E_{JT}S^2)$, $\tilde{t}=4t/3$.
$C_{i,j} \equiv \cos \left[(\Theta_i + \Psi_{ij})/2 \right]$ with
$\Psi_{i,i+x} = \Psi_{i+x,i} = + \pi/3$, $\Psi_{i,i+y} =
\Psi_{i+y, i} = - \pi/3$ and $\Psi_{i,i+z} = \Psi_{i+z,i} = \pi$.

The effective Hamiltonian of eq. (\ref{effectiveE}) is a
\textit{classical} spin-charge-orbital model with no quantum
fluctuations. If the charges are assigned specific positions (so
as to minimise the energy), the model reduces to a spin-orbital
model. It is different from the spin-orbital model proposed for
undoped LaMnO$_3$ obtained by projecting out double
occupancies\cite{spinorbital} because double occupancy is
irrelevant in the limit being explored here. The orbital (and JT
distortion orientation) variables on neighbouring sites are not
directly coupled in this model (as  $C_{ij}$ involves only one
orbital angle $\Theta_i$). Such a coupling would arise if one
takes into account a direct coupling between JT distortions on
neighbouring sites, as in the cooperative JT model. Nevertheless,
even in this simplified approach, the virtual double exchange
lifts the degeneracy between different orientations of the JT
distortions and the corresponding orbital degeneracy. In addition,
it clearly favours ferromagnetic bonds and charge
disproportionation.

\begin{figure}[htbp]
\centerline{ \psfig{file=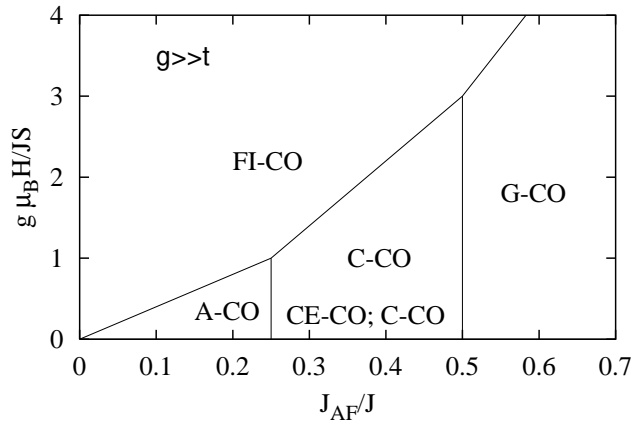,width=6cm,angle=-90}}
\caption{Phase diagram when $g/t \gg 1$ at $T=0$ ($x=0.5$). The
phases are the same as in Fig. \ref{pd-f}.
$JS^2=\tilde{t}^2/2E_{JT}$. The CE-CO and C-CO phases are
degenerate at zero-field, but the latter wins at finite fields.} \label{pdgt-f}
\end{figure}

We can now understand the strong-coupling limit of the phase
diagram (Fig. \ref{pd-f}), and furthermore even in the presence of
a magnetic field, by estimating the energies of the various phases
using equation (\ref{effectiveE}). For the fully charge
disproportionated phases the energies per site are obtained by
minimising $\tilde{\cal H}$ with respect to the canting angle (at
finite fields) and orbital angle (the latter depends on the field
for the CE phase, and only the leading term in $H \rightarrow 0$
is given):

\begin{eqnarray}
E_{FI-CO} &=& 3J_{AF}S^2-3JS^2/2 - \mathrm{g}\mu_B HS\nonumber \\
E_{A-CO} &=& J_{AF}S^2-3JS^2/2 -\frac{(\mathrm{g}\mu_B H)^2}{8J_{AF}}\nonumber \\
E_{C-CO} &=& - J_{AF}S^2-JS^2 -\frac{(\mathrm{g}\mu_B H)^2}{16J_{AF}-2J} \nonumber \\
E_{CE-CO} &=& - J_{AF}S^2-JS^2 -\frac{(\mathrm{g}\mu_B H)^2}{16J_{AF}-J} \nonumber \hspace{0.3cm} (H \rightarrow 0) \\
E_{G-CO} &=&   -3 J_{AF}S^2 -\frac{(\mathrm{g}\mu_B H)^2}{24J_{AF}-6J} \nonumber
\end{eqnarray}
At zero field, there is only one free parameter, $J_{AF}/J$, which
determines the relative energies. At $J_{AF}=0$, the FI phase
(which is orbitally disordered in this limit) is degenerate with
the A-type phase (with $(x^2-y^2)$ orbital order), but the latter
is favoured as soon as $J_{AF}>0$.  There is a succession of {\it
first-order} phase transitions as $J_{AF}/J$ is increased, first
to the CE-CO phase (degenerate with the C-CO phase) at
$J_{AF}/J=1/4$ and then to the G-CO phase (see Fig. \ref{pdgt-f}).
In terms of the original variables, the first transition at $1/4$
is located at $J_{AF}S^2/t=4tK/(9g^2)$; the second transition is
at $J_{AF}S^2/t=8tK/(9g^2)$.  The phase boundaries given by these
equations in the strong-coupling regime are displayed as
dash-dotted lines in the phase diagram (Fig.  \ref{pd-f}) and are
in good agreement with the phase boundaries obtained numerically.

Similarly we can compare the energies of the different phases as a
function of the magnetic-field and draw the corresponding phase
diagram in the $(J_{AF}/J$, $\mathrm{g}\mu_B H/JS$) plane
(Fig. \ref{pdgt-f}). The degeneracy between the C and the CE phases is
lifted and the C phase wins at finite fields. This is because the
fourth term of (\ref{effectiveE}) favours Wigner-crystal type of
ordering.  For the C phase, for instance, the critical field is
$\mathrm{g}\mu_B H_c=8J_{AF}S-JS$.

As noted earlier, in the limit of large $g/t$ much of the physics
is insensitive to the inclusion in the model of the on-site
Coulomb interaction between different orbital states, $Un_{i
\alpha} n_{i \bar{\alpha}}$. The total energy is hardly affected
since double occupancy is much reduced. We emphasise again that
this is contrary to what happens in the other limit $g/t \ll 1$
where the electron density is uniform. In the latter case, it is
known that $U$ by itself will induce charge-ordering in the CE
phase,\cite{Khomskii} at least if the latter is not destabilised
by other phases.\cite{Shen} In the CE phase at large $g/t$, $U$
will slightly modify the charge contrast by pushing the electrons
further off the corner sites. We have performed a self-consistent
Hartree-Fock calculation to confirm this point. At $g=0$, the
calculation gives the same results as the slave-boson
approach.\cite{Khomskii} At $g/t=7$ and for the optimised lattice
distortions, the charge contrast increases, with respect to $U=0$,
by a very small amount of order 0.05 for $U/t$ even as large as
$25$. Therefore, we conclude that the effect of $U$ is small and
does not change the nature of the insulating phases in the limit
of large $g/t$.

We next consider the interesting  question as to what the
appropriate low energy effective Hamiltonian replacing
(\ref{effectiveE}) is when $t/E_{JT}$ becomes sufficiently large
that perturbation theory in $t/E_{JT}$ and the Hamiltonian
(\ref{effectiveE}) are not valid anymore. We argue below, by
studying the excitations and instabilities of the original model
(\ref{Hamiltonian}) in the ferromagnetic phase, that the effective
model that replaces (\ref{effectiveE}) when $t/E_{JT}$ gets larger
takes a similar form except that mobile electrons have to be
included.

\section{Instabilities of the Periodic Phases}
\label{Instabilities-s}

We have discussed in the previous section the various phases stable in
the thermodynamic limit that are periodic with a 8-sublattice
unit-cell. We will discuss in this section several instabilities that
point to non-periodic phases, at half-doping
(\ref{InstabilityFerro-ss}) and also upon doping (\ref{doping-ss}) or
addition of a magnetic-field (\ref{mag-f-ss}).  We will show that the
ferromagnetic insulating phase (FI-CO) is in fact unstable when $g/t$
is lowered below a critical value $g_c / t \sim 6.8 $.  This
instability occurs {\it before} any of the transitions discussed above
(at $g_c / t \sim 6.3$ and $ 5.9$) take place.

For this purpose, we study the energetics of defects in the lattice
distortion pattern of the periodic phases. We consider particle and
hole excitations {\it accompanied by single site JT defects}. We
consider both types of defects, one where we add a distortion on a
site that was previously undistorted, and the other where we remove
the distortion of a site that was distorted. Without the lattice
distortion defect, the lowest energies of the particle or hole
excitations accessible are the appropriate gaps determined by the
band-structures discussed in section \ref{results-ss}
(Fig. \ref{chargegap-f}). The defect modifies locally the JT energy
level, and hence constitutes a scattering potential for the particle
and hole excitations. The problem lacks lattice translation
invariance, and we have solved it by exact numerical diagonalisation
of Hamiltonian (\ref{Hamiltonian}) \textit{represented in
real-space}. We consider a problem of $N$ sites (up to $N=1728$) with
one special site, and we calculate all the eigenvalues and the total
energy.  A key question is whether bound states with energies lower
than that allowed by band-structure can appear near the defect. We
find that they do in several cases, and when their binding energy
exceeds the gap, it signals an instability of the periodic phase.

\subsection{Instability of the Ferromagnetic Insulating Phase when $g/t$ is decreased}
\label{InstabilityFerro-ss}

We consider first the FI-CO phase at half-doping (pictured in Fig.
\ref{pd1-f}). It is stable for very small $J_{AF}$ and large $g/t$
(see the phase diagram in Fig. \ref{pd-f}). Out of the two sites in
the unit-cell, one site is distorted with a distortion orientation
that favours the $d_{x^2-y^2}$ orbital.

We now consider the problem when one introduces a single site JT
defect: the amplitude of the distortion $Q$ of the FI-CO phase is
maintained at $N/2-1$ sites except at one site where the distortion is
now reduced to $Q-Q_d$. $Q_d$ takes all values from $0$ (no defect) to
$Q$ (the lattice distortion has been completely removed on this
site). The excess energy of such a state is given by:

\begin{eqnarray}
E-E_0 = E_{el}(N_0,Q,Q_d) - E_{el}(N_0,Q,0) \nonumber \\  +
\frac{1}{2}K(Q-Q_d)^2 - \frac{1}{2}K Q^2 \label{ebs}
\end{eqnarray}
Here $E_0$ and $E_{el}(N_0,Q,0)$ [$N_0=N/2$ is the number of
electrons] are the total and {\it electronic} ground state energies of
the optimal FI distorted phase, obtained as a function of $g/t$ by
minimising with respect to $Q$ as discussed in the previous
section. $E_{el}(N_0,Q,Q_d)$ is the {\it electronic} ground state
energy of the defective state. One expects a gain in lattice energy
and a loss in electronic energy,
because one energy level has been raised at one site. To evaluate the
latter, we first solve numerically the problem of the one electron
eigenvalues in the presence of the extra single site potential for a
finite-size system.  Then we calculate $E_{el}(N_0,Q,Q_d)$ by filling
the $N/2$ lowest one-electron levels. We have so far
considered 3d systems with up to $N=1728$ sites.

\begin{figure}[htbp]
\centerline{\psfig{file=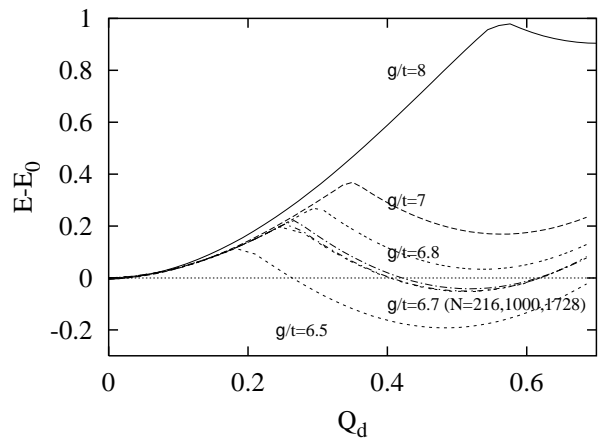,width=6.0cm,angle=-90}}
\caption{Energy change when a single JT defect is introduced in the
FI-CO phase, vs. $Q_d$. $Q-Q_d$ is the JT distortion on a defect site; all
the other occupied sites having the same distortion $Q$. We see that
there is an excitation with $Q_d \sim Q$ that softens when $g/t$
decreases. The excitation corresponds to a band particle-hole
excitation with the removal of a lattice distortion of one site, while
the $Q_d=0$ minimum is the polaron. The softening for $g_c/t \sim 6.8$
signals a phase transition with proliferation of mobile electrons and
defects.  Finite-size effects are small and shown for $g/t=6.7$
($N=216,1000,1728$).} \label{deltaelastic-f}
\end{figure}

Fig. \ref{deltaelastic-f} shows the energy $E-E_0$ plotted vs.
$Q_d$ for different values of $g/t$. We have checked that finite
size effects are negligible (the curves corresponding to $N=216,
1000, 1728$ are given for $g/t=6.7$ in Fig. \ref{deltaelastic-f}).
For large $g/t$, the energy is positive but there is a local
minimum at large $Q_d$ which can be described as a particle-hole
excitation with reduced distortion on one site. When $g/t$
decreases, this excitation softens and vanishes at $g_c/t \sim
6.8$. We believe that this signals the onset of a new phase where
such defects are energetically favourable and thus proliferate in
the system.

From the calculation of the energy levels in the presence of the
defect, we find that there is no bound state within the gap for the
$Q_d$ that minimises the energy. The electron occupies a higher-energy
band-like state and is mobile.  The instability therefore corresponds
to the energy of this mobile electron crossing the chemical potential
(i.e., $-E_{JT}$). This suggests that the proliferation of the defects
leads to the conversion of some small fraction of the localised
electrons into mobile electrons moving on weakly distorted sites,
resulting in a metallic phase. Such a state would not be accessible in
the minimisation procedure of section \ref{pd-s} (which has a maximal
unit cell of 8 sites) even if the defect sites were to arrange
themselves in a super-lattice.

The situation is rather similar to that described by Ramakrishnan
\textit{et al.} at the metal-insulator transition in hole doped
manganites in the orbital-liquid regime;\cite{Ram} except that we have
here a calculation in the context of a microscopic model that
explicitly suggests such a picture even when $g/t$ is not very
large. Thus we can identify the high-energy mobile electrons as the
\textit{broad-band} $b$-like electrons of ref. [\onlinecite{Ram}], and
the low-energy localised states as the $\ell$ polarons. In the present
context, at half-doping and above the
transition ($g>g_c$), all the electrons occupy the $\ell$ states,
which form a regular checker-board array (Fig. \ref{pd1-f}). The sites
are singly occupied and $U$ does not play a crucial role. This is no
longer the case below the transition when we start to transfer some
electrons from the $\ell$ states to the $b$ states. The $b$ states are
delocalised over the empty sites but also visit the sites occupied
with $\ell$ electrons. Double occupancies become inevitable and $U$
has to be taken into account in order to determine accurately the
properties of the metallic state. The question of what kind of new
metallic state arises for JT couplings just below the instability is
clearly interesting. The mobile electrons, for instance, may be able
to destroy the orbital and charge order. While a study of such issues
is beyond the scope of the present article, the above results 
suggest than one should add mobile electrons to the strong-coupling Hamiltonian
(\ref{effectiveE}), in order to describe metallic
phases with possible partial orbital/charge order:

\begin{eqnarray}
&&  \tilde{\cal H} = - E_{JT} \sum_i n_{\ell i}  + \sum_{<ij>}
J_{AF} \textbf{\mbox{S}}_i . \textbf{\mbox{S}}_j \nonumber \\ && -
{\frac {J}{2}} \sum_{<i,j>} (\textbf{\mbox{S}}_i \cdot
\textbf{\mbox{S}}_j+S^2)\left[ n_{\ell i}(1-n_j)
C_{i,j}^2  + ( i \leftrightarrow j)\right] \nonumber\\
&& - \sum_{<i,j> } t_{ij}^{\alpha \beta} b_{i \alpha }^{\dagger}
b_{j \beta } + \sum_i U n_{i \alpha}^b n_{i\bar{\alpha}}^{\ell}
\label{effectiveEmobile}
\end{eqnarray}
where the orbital index $\alpha$ of the mobile $b$ electrons takes
both values on the undistorted sites, but is constrained to be
orthogonal to the $\ell$ orbital on the occupied sites, and the
other quantities have the same meanings as in eq.
(\ref{effectiveE}). For infinite $U$ the mobile electrons can-not
hop to the $\ell$ sites at all, and the last pair of terms can
simply be replaced by $- \sum_{<i,j> } t_{ij}^{\alpha \beta} b_{i
\alpha}^{\dagger}  b_{j \beta}(1- n_{\ell i})(1- n_{\ell j})$.
This Hamiltonian needs to be studied in a framework that can
handle the strong interaction effects, such as the dynamical
mean-field theory, in a similar way as was done before for the
orbital-liquid state.\cite{Ram} Note furthermore that the above
Hamiltonian does not include $\ell-b$ hybridisation effects, which
must be included in order to describe properties sensitive to
$\ell-b$ coherence which can develop at sufficiently low
temperatures in the metallic phases.\cite{Ram} It is
straightforward to generalise the Hamiltonian to include these
effects, as well as cooperative JT effects. 

\subsection{Instability of the CE Phase upon Doping}
\label{doping-ss}

In the band picture of the CE phase\cite{Terakura,Khomskii},
doping with electrons, corresponding to $x < 1/2$ (resp. holes,
corresponding to $x > 1/2$) provides mobile carriers in the
conduction (resp. valence) band. In either case, the system will
be metallic. This is contrary to experiment, where, in most cases,
the system remains insulating for $x>1/2$, but typically becomes
metallic quickly for $x<1/2$. The transition to the ferromagnetic
metal for $x<1/2$ has been described as being due to the crossing
of the energies of the CE and ferromagnetic metallic
states.\cite{Khomskii} The transition is then naturally
first-order. However, as discussed in ref. [\onlinecite{Ram}] even
for $x<1/2$ a simple band picture of the ferromagnetic metallic
state is severely limited. Apart from that, the band picture fails
to describe the insulating character of the regime $x>1/2$ and the
particle-hole asymmetry around $x=1/2$. We discuss this issue
next.

It was pointed out a long time ago by de Gennes\cite{deGennes}, in
the context of slightly doped LaMnO$_3$, that adding carriers to
the antiferromagnetic phase of LaMnO$_3$ may favour canted
structures. The qualitative argument is that at small
concentration the carriers gain kinetic energy which is linear in
the canting angle whereas the loss of magnetic energy is quadratic
in the canting angle. By the same token, adding carriers to the CE
phase should lead to canting of the core-spins. As such phases
interpolate between the CE and FM phases, the transition to
ferromagnetism should be naively second-order.

In view of this, we have calculated the energy of homogeneous CE
canted phases (defined in Fig. \ref{latticePertu-f}) for different
carrier concentrations on either side of $x=1/2$ (i.e., retaining
the 8-sublattice periodic structure even when $x \neq 1/2$).  We
find that canting is favourable for adding electrons to the
half-doped system but not for adding holes, as de Gennes's general
argument is valid only for very small carrier concentration and
breaks down quickly on the hole side, due to special features of
the CE state. We have calculated, in addition, the energy in the
presence of a single-site defect in the JT distortion as in the
previous subsection. We find that, when $g/t$ is sizeable, canting
is in competition with self-trapping of the carriers in JT
defects. We find that the energy gain due to trapping is linear in
the carrier concentration (and thus dominates at low
concentration) whereas it is quadratic for the canting. For
intermediate values of $g/t$, this results in a first-order
transition to a canted metallic phase when electrons are added to
the half-doped system (i.e., for $x < 1/2$), and the persistence
of a CE-type phase with self-trapped carriers when holes are added
(i.e., for $x > 1/2$).

\subsubsection{Canted Phases}
\label{doping-ss-canted}

\begin{figure}[htbp]
\centerline{ \psfig{file=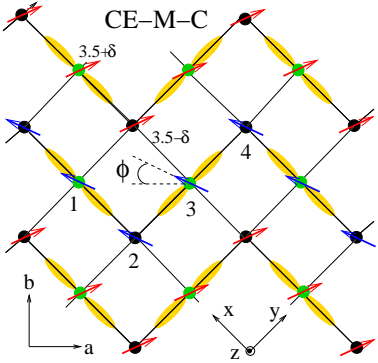,width=5cm,angle=-0}}
\caption{(color online). Canted CE phase with canting angle $\phi$
(CE-M-C). The electrons can now hop from chain to chain. The original
1d zig-zag chains are marked with thicker lines.}
\label{latticePertu-f}
\end{figure}

\begin{figure}[htbp]
\psfig{file=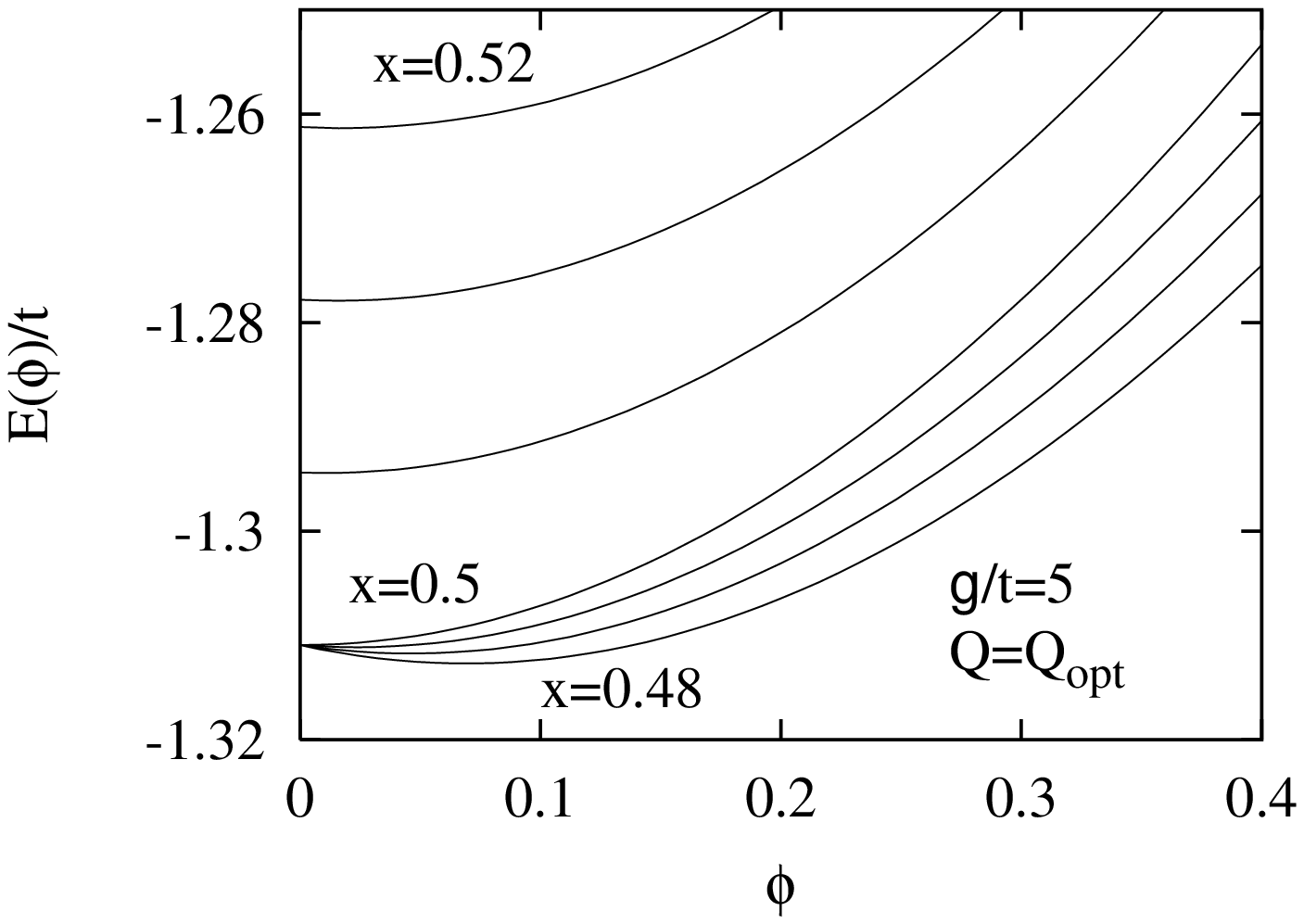,width=6cm,angle=-90} \\
\psfig{file=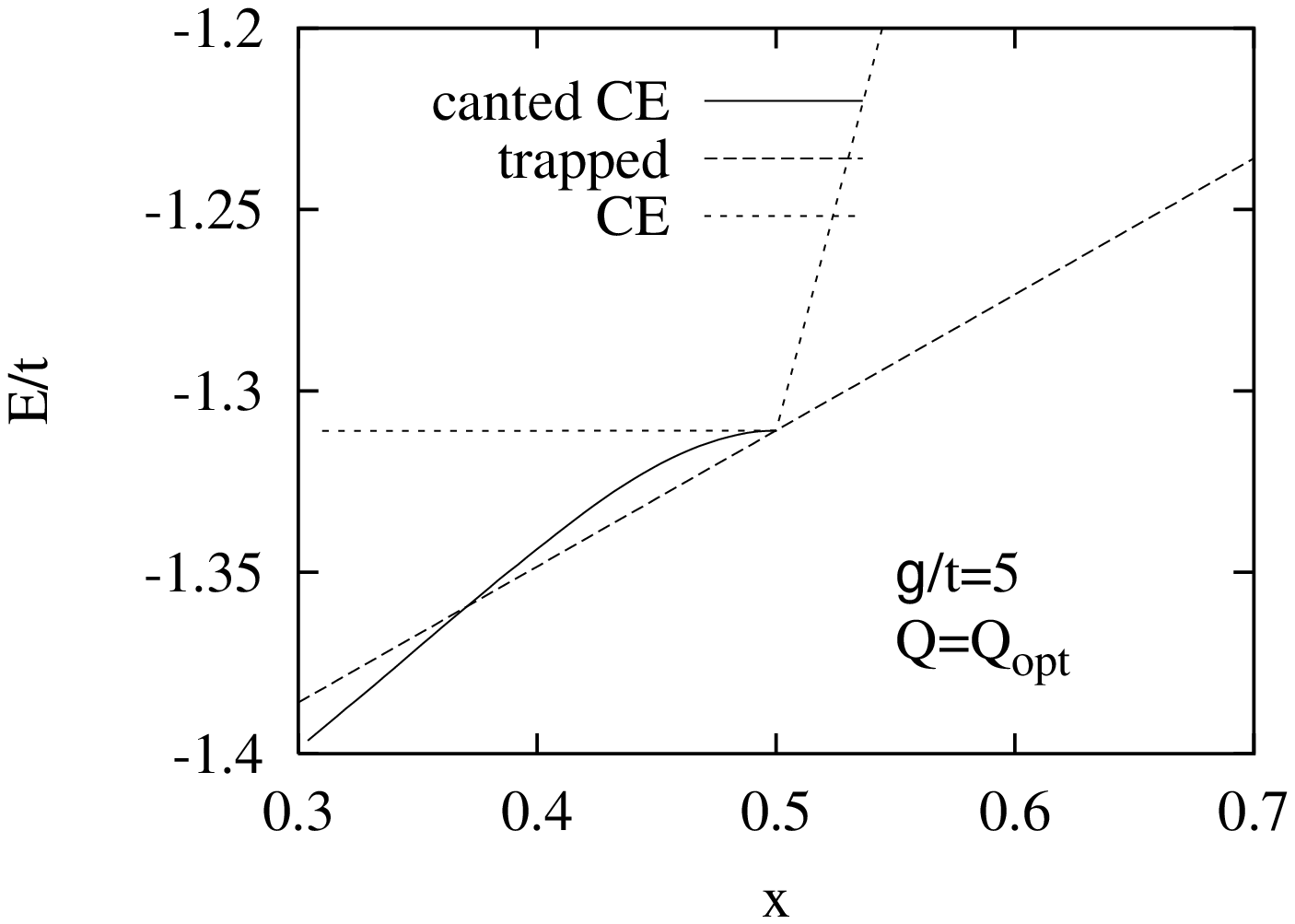,width=6cm,angle=-90} \caption{Top.
Energies of the canted phases as function of the canting angle,
for different concentrations ($g/t=5$ and $Q=Q_{opt}=0.35$). For
$x<0.5$, there is a finite angle that minimises the energy. For
$x>0.5$, the angle is zero and the CE state is stable. Bottom:
Comparison of the energies of the different phases. The canting
angle is chosen such as to minimise the energy as in the top
figure.} \label{energiesdoping-f}
\end{figure}

We first consider the total energy of a homogeneous canted CE
phase as a function of the canting angle $\phi$ (see Fig.
\ref{latticePertu-f} for definition) for different values of
doping (Fig. \ref{energiesdoping-f}, top). We take the JT
distortions that optimise the energy (as discussed in section
\ref{pd-s}), except that we neglect the small distortions on the
corner sites for simplicity.\cite{remark} When adding electrons
($x<0.5$), the energy of the system is lowered by a finite canting
angle (Fig. \ref{energiesdoping-f}, top). The higher the
concentration the higher the canting angle. When adding holes
($x>0.5$), however, it turns out that the system prefers $\phi=0$.

\begin{figure}[htbp]
\centerline{
\psfig{file=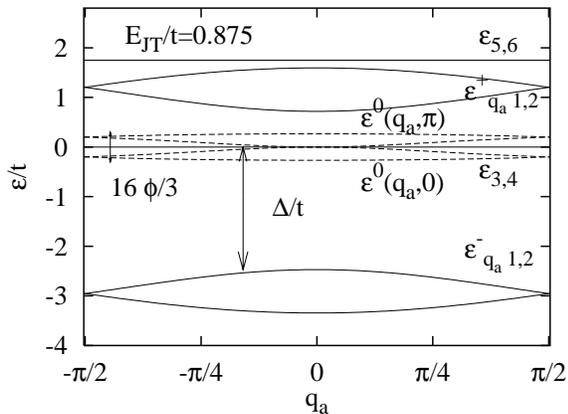,width=6cm,angle=-90}}
\caption{Band structure of the 1d zig-zag chains (solid lines) with
Jahn-Teller distortions ($E_{JT}/t=0.875$), given by eqs.
(\ref{bseq1}), (\ref{bseq10}) and
(\ref{bseq2}).\cite{Terakura,Khomskii,Popovic} At $x=1/2$, only the
lowest band is filled. Also shown is the splitting of the zero-energy
states (dashed lines) when the core spins are tilted away by an angle
$\phi$ (whence their dispersion becomes 2-dimensional).}
\label{bandstructure-f}
\end{figure}

We can understand these results by considering in more detail the
band structure of the CE phase.\cite{Terakura,Khomskii,Popovic} As
discussed earlier for $x=1/2$ (section \ref{pd-s}), the bridge
sites are orbitally ordered in the $(3x^2-r^2)/(3y^2-r^2)$
pattern.\cite{Khomskii} If the Jahn-Teller distortions occur in
such a way as to favour precisely the alternation of
$(3x^2-r^2)/(3y^2-r^2)$ orbitals, the energy of the system is
further lowered. The band structure has four dispersive bands and
four non-dispersive bands (two at zero and two at finite energy),
as shown in Fig. \ref{bandstructure-f} (see Refs.
[\onlinecite{Khomskii,Terakura,Popovic}])and described by:

\begin{eqnarray}
\epsilon_{q_a 1}^{\pm} &=& -E_{JT} \pm \sqrt{E_{JT}^2 + \tilde{t}^2 (2+\cos q_a)} \label{bseq1} \\
\epsilon_{q_a 2}^{\pm} &=& -E_{JT} \pm \sqrt{E_{JT}^2 + \tilde{t}^2 (2-\cos q_a)} \label{bseq10} \\
\epsilon_{3,4} &=& 0, \hspace{1cm} \epsilon_{5,6}=2E_{JT} \label{bseq2}
\end{eqnarray}
where $\tilde{t}=4t/3$ and $E_{JT}=gQ/2$, and the wave-vector $q_a$,
which takes values in the reduced Brillouin zone $[-\pi/2, \pi/2]$, is
parallel to the chain direction. The band structure is analogous to
that of Ref. [\onlinecite{Khomskii,Terakura}] with a charge ordering
coming from the Jahn-Teller distortions, as in Ref.
[\onlinecite{Popovic}], but determined self-consistently. At $x=1/2$,
the lowest band is completely filled.

The zero energy band is made up entirely of the states from the
corner sites (see appendix A).  The charge gap, from the top of
the filled valence bands to the zero energy states, is given by:

\begin{equation}
\Delta = E_{JT} + \sqrt{E_{JT}^2+\tilde{t}^2} \label{gap}
\end{equation}
When the core spins are canted away from $\phi=0$, the degeneracy
of the zero energy states of the zig-zag chains is lifted, and
they form bands which disperse. {\it To first order in the canting
angle}, the dispersion arises from the coupling of the
$(3z^2-r^2)$ orbitals at the corner sites along the $z$-direction,
and is given by:

\begin{equation}
\epsilon^0(q_a,q_z)= -4t \phi \frac{1+\cos q_a}{2+\cos q_a} \cos q_z
\label{dispersionzeroband}
\end{equation}

Added electrons (with respect to the reference state $x=1/2$) occupy
the bottom of this new band (properly folded in the reduced Brillouin
zone, Fig.  \ref{bandstructure-f}) at $q,q_z \sim 0$. Each electron
therefore gains an energy $-{\beta}_e \phi t$ with ${\beta}_e=8/3$,
which is completely independent of $g$. On the other hand, there is an
energy loss $\kappa \phi^2$ per site, where $\kappa$ is the effective
spin stiffness towards canting.  There are two contributions to
$\kappa$. One comes from the direct super-exchange. In the CE phase
with $\phi=0$, two of the neighbour spins are parallel and the other
four neighbour spins are antiparallel. This gives a contribution
$4J_{AF} S^2 \phi^2 $ to the energy. There is also a contribution from
the double exchange, which we calculate numerically by calculating the
change in the total kinetic energy as function of $\phi$. From this we
extract a stiffness in the limit of small $\phi$,
$-{\kappa}_e=f(g/t,\delta c)$ where $\delta c \equiv 1/2-x$ is the
filling fraction in excess of that at half-doping, so that
$\kappa=4J_{AF}S^2-{\kappa}_e$.

The canting angle is then given by minimising the excess energy
per site, $E-E_0 = \kappa \phi^2 - {\beta}_e t \delta c \phi$, and
is given by $\phi= {\beta}_e t \delta c/2\kappa$.  Note that the
linear dependence in $\delta c$ is valid only for sufficiently
small $ \delta c$ whence the additional electrons occupy states
near the bottom of the band. There are in fact $N/8$ states with
energy smaller than 0 and $N/8$ states with energy greater. So the
linear dependence is expected to be reasonable for $\delta c \ll
1/8$. The total minimised excess energy is then

\begin{equation}
E-E_0 = -\frac{\beta_e^2 t^2}{4 \kappa} (\delta c)^2,
\label{excess_energy_canted}
\end{equation}
quadratic in the electron concentration ($E_0$ is the energy of
the CE phase at $x=1/2$). The canted CE phase thus has a lower
energy than that of the CE phase by this amount, whenever $\delta
c > 0$. This explains the quadratic behaviour in $\delta c$ found
numerically in the previous paragraph (Fig.
\ref{energiesdoping-f}, bottom). When $\delta c$ increases
further, the canting angle increases until eventually the system
becomes fully ferromagnetic via a second-order phase transition.
Within the above picture, as soon as $\delta c > 0$, the system is
metallic because the additional electrons occupy the dispersing
conduction band. But note that, since the dispersion is mainly
along the z and the chain directions to leading order in the
canting angle, the metalicity generated by such a mechanism would
be highly anisotropic.

\begin{figure}[htbp]
\centerline{
\psfig{file=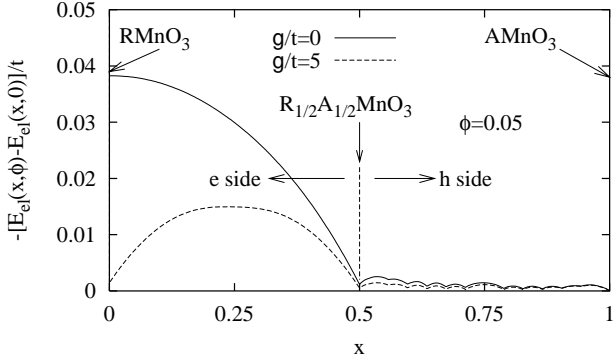,width=5cm,angle=-90}}
\caption{Energy gained by canting of the CE phase for a fixed
canting angle of $\phi = 0.05$ as a function of doping away from
half doping, showing the particle-hole asymmetry.}
\label{en-vs-x-fix-phi}
\end{figure}

Surprisingly, the situation turns out very differently  when
excess holes are added to the half-doped system. Naively, one
might have thought that canting the spins will push up the states
at the top of the lower band and that the added holes will thus
gain energy, similarly to the case of electrons. But according to
the numerical results of Fig. \ref{energiesdoping-f}, this is not
what happens. The energy is minimal at $\phi=0$, i.e. when the
spins are \textit{not} canted. The reason is that the energy gain
for the whole system is not linear in the number of additional
holes, except for extremely small $\delta c$. This is even more
emphatically and dramatically evident from Fig.
\ref{en-vs-x-fix-phi}, which shows that the energy gained by
canting for a small fixed canting angle ($\phi=0.05$) as a
function of doping, is extremely particle-hole asymmetric.
While the gain is indeed linear in the deviation from half doping
on either side for the smallest values of $|\delta c|$, it drops
quickly and substantially below this, and stays small for $x>1/2$,
i.e., for added holes. In contrast, when electrons are added, the
energy gain from canting remains large for large $|\delta x |$.
(We emphasise that the above results are only for the canted CE
state for any $x$; hence they are most meaningful near $x= 1/2$.)

The reason for this asymmetry is the difference in the dispersion
of the electron and hole bands in the canted CE phase. As
discussed in detail above, the conduction band in the canted CE
phase (into which electrons get added) disperses only in two
directions, and hence the corresponding density of states is
constant, over a bandwidth of order $t \phi$. Hence the energy
gain from canting remains substantial up to half occupancy of the
conduction band, corresponding to $\delta c = 0.25$, beyond which
the energy gain starts reducing because of cancellations, as is
clear from Fig. \ref{en-vs-x-fix-phi} (dashed line). In contrast,
the valence band in the canted CE phase which accommodates the
holes (and which we have not discussed in detail), disperses in
all three directions, resulting in a density of states which
starts from zero at the top of the band. Typically when the change
in the chemical potential is smaller than $t \phi$, the additional
holes occupy the higher energy states of the new band and the
energy gain is linear in the number of additional carriers. On the
other hand, when the chemical potential is of order $t \phi$, both
sides of the new band are occupied and there is no energy gain.
The condition is expressed as $\delta c < t \phi \rho(\epsilon_F)$
where $\rho(\epsilon_F)$ is the density of states at the Fermi
energy for the system with the concentration $c+\delta c$. When
the density of states is small, the range within which the energy
gain is linear in the number of carriers is small. In this range
and out of it, the canting angle is then very small because it
does not scale with the number of carriers anymore, as is
confirmed by Fig. \ref{en-vs-x-fix-phi}.

\subsubsection{Trapping of Added Carriers in the CE Phase}
\label{doping-ss-trapped}

\begin{figure}[htbp]
\psfig{file=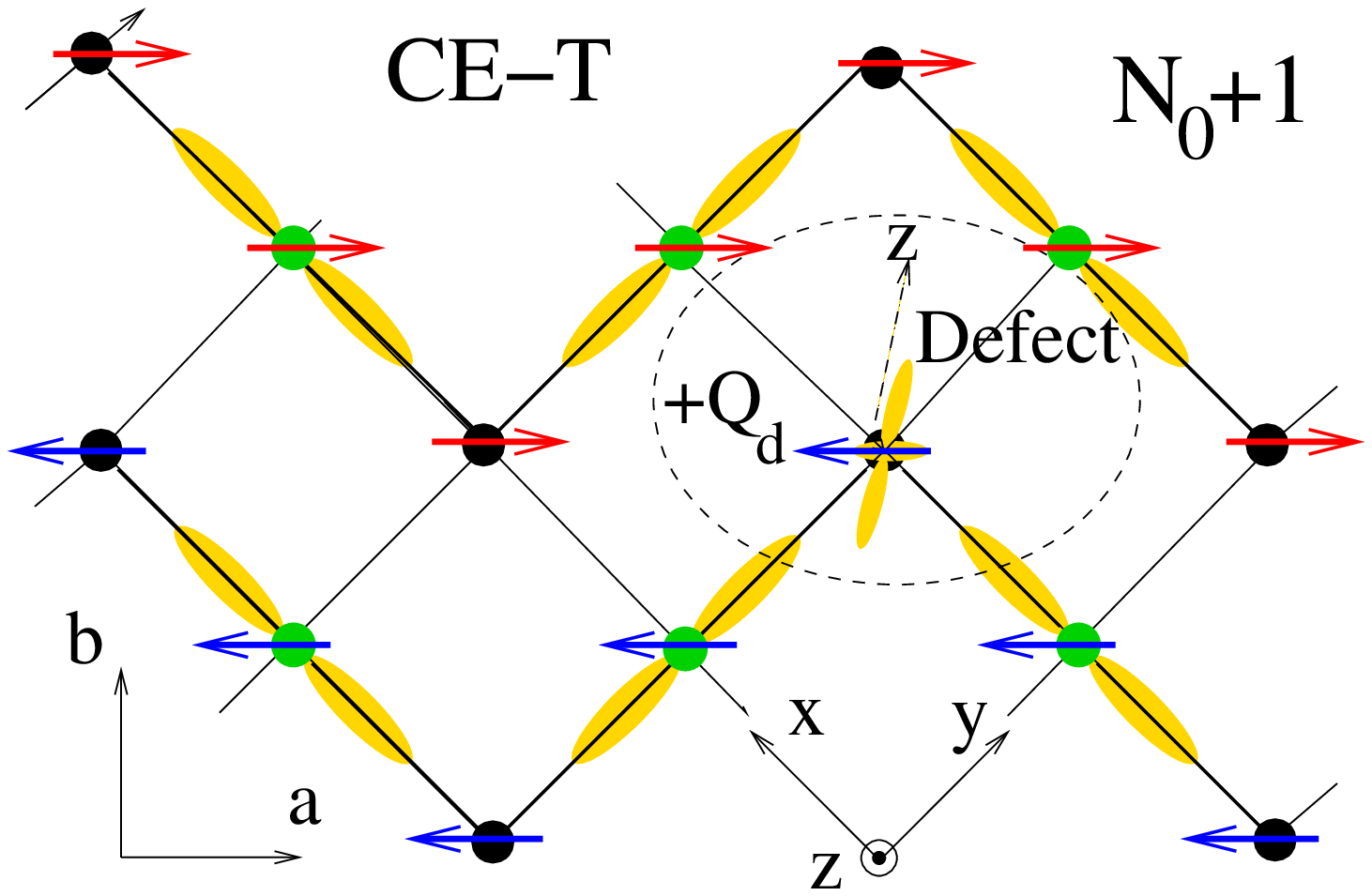,width=5cm,angle=-0}
\centerline{
\psfig{file=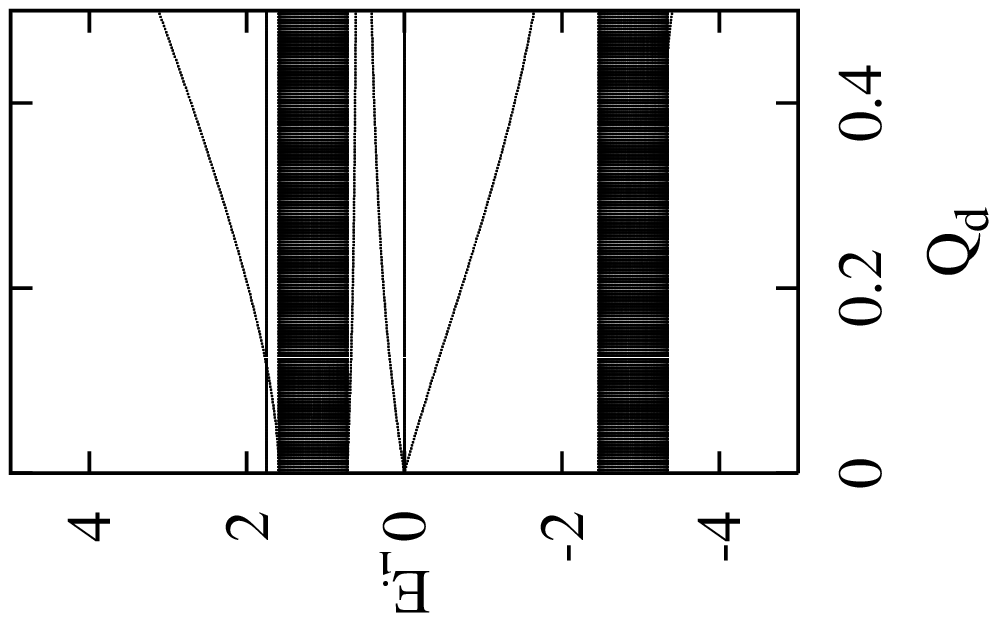,width=5cm,angle=-90}
\hspace{-0.7cm}
\psfig{file=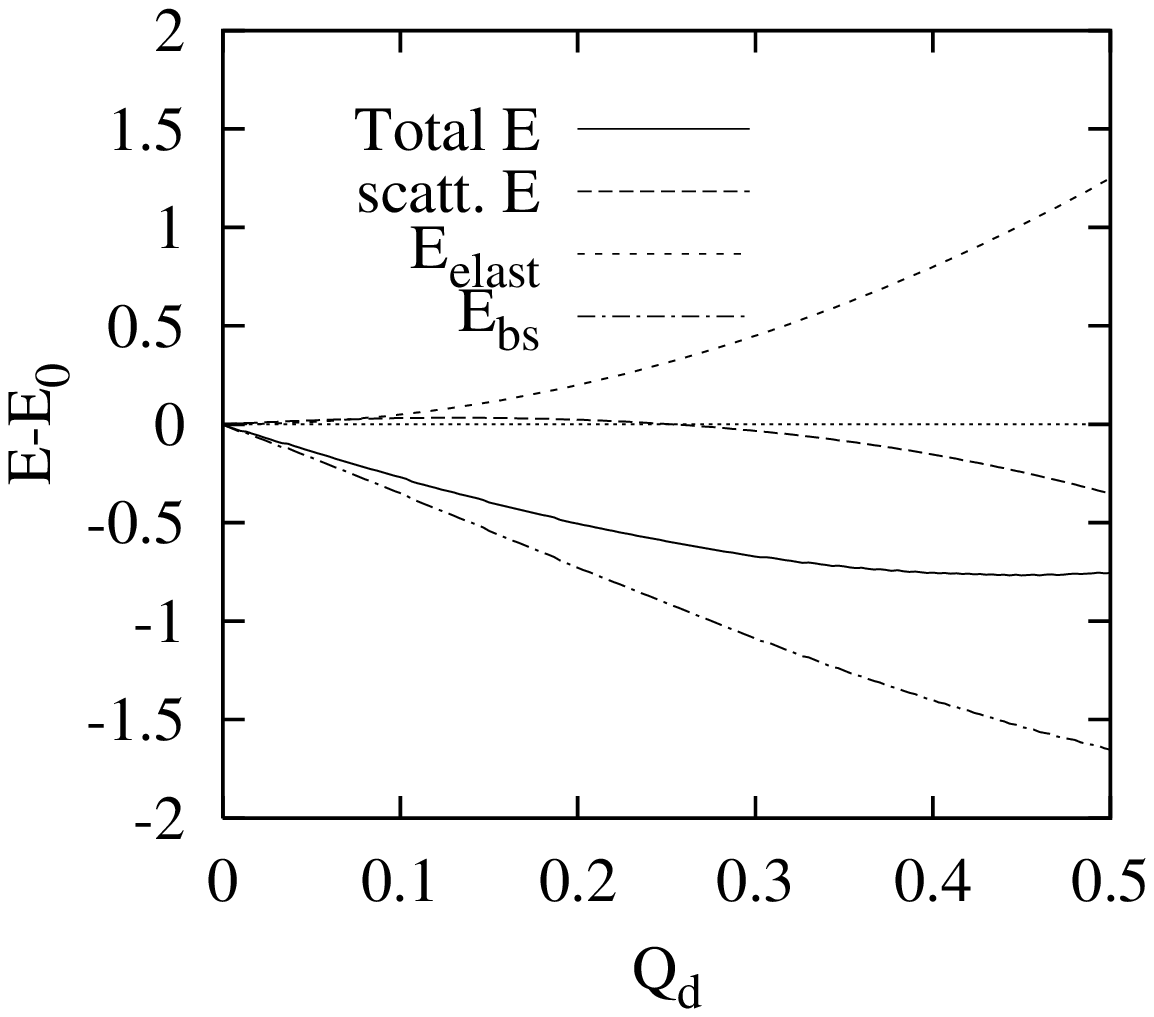,width=5cm,angle=-90}}
\caption{(color online). Definition of the defect (top), that is an
additional distortion of strength $Q_d$ on a single corner site ($(3z^2-r^2)$
orbital). Energy levels vs. $Q_d$ (bottom left). At
$Q_d=0$, there is no defect and the band structure is that of
Fig. \ref{bandstructure-f}. At $Q_d>0$, bound states appear within the
gap. Total excess energy of the system with $N_0+1$ electrons (bottom
right). This can be seen as the sum of the energy of the bound state
$E_{bs}$, the elastic energy $E_{elast}$ and defines the scattering
energy $E_{scatt}$. $g/t=5$, $Q=Q_{opt}=0.35$ ($E_{JT}=0.875$).}
\label{levelsdefect-f}
\end{figure}

\textit{Trapping of Added Electrons:} We next explore the competition
between the above process and the possible trapping of the added
electrons in lattice distortions by allowing for an additional lattice
distortion on one special site and $N_0+1$ electrons ($N_0=N/2$).  In
other words we look at the cost or gain of energy involved in trapping
one additional electron in the CE state. Since the added electron goes
into a band made up from states belonging to the corner sites which is
originally undistorted, we choose the special site with the added
distortion to be one of the corner sites, and of such an orientation
as to lower the energy of the $(3z^2-r^2)$ orbital
(Fig. \ref{levelsdefect-f}). As in \ref{InstabilityFerro-ss},
translation invariance is now broken, and we find the one-electron
energy eigenvalues by diagonalizing exactly the problem on large
lattices, as a function of the strength of the additional distortion,
$Q_d$. We show the energy levels in Fig. \ref{levelsdefect-f},
left. In addition to the band states (black areas), we also find a
couple of bound-states within the gap (Fig. \ref{levelsdefect-f},
left).

The total excess energy including the cost of elastic energy to
create such a defect in the lattice is calculated by filling up
the energy levels with $N_0+1$ electrons:

\begin{equation}
E - E_0 = E_{el}(N_0+1,Q_d) -  E_{el}(N_0+1,0) + \frac{1}{2} K Q_d^2
\end{equation}
The energy can be viewed as the sum of three different
contributions, each of which is separately shown in Fig.
\ref{levelsdefect-f}: there is the electronic energy gain for the
electron bound to the defect, the scattering energy for the
electrons that are scattered by the defect (both of which are
contained in $E_{el}(N_0+1,Q_d)$) and the elastic energy cost for
creating such a distortion.  When all are put together, it turns
out that it is favourable to create a defect with an energy gain,
$\tilde{E}^e_{JT}$.  For instance, in Fig. \ref{levelsdefect-f},
when $g/t=5$, the energy gain is $\tilde{E}^e_{JT}= 0.75t \sim
0.86E_{JT}$, slightly smaller than $E_{JT}$ due to band structure
effects. The problem can not, indeed, be reduced to polaron
formation purely with the states of the conduction band by
ignoring the valence band (that would give $\tilde{E}^e_{JT}=
E_{JT}$). The scattering of electrons by the defect and the level
repulsion between the two bands (Fig. \ref{levelsdefect-f})
accounts for the reduced value of $\tilde{E}^e_{JT}$ found above.

For a small concentration of additional electrons, $\delta c$, the
energy gain with respect to the CE phase, given by

\begin{equation}
E-E_0=-\tilde{E}^e_{JT} \delta c, \label{excess_energy_CEtrapped}
\end{equation}
is linear with the concentration of additional electrons. Here
$E_0$ is the energy of the perfect CE phase with additional
electrons occupying the (undistorted) zero-energy band states.

When we compare eq. (\ref{excess_energy_CEtrapped}) and eq.
(\ref{excess_energy_canted}), it is clear that the linear dependence
(\ref{excess_energy_CEtrapped}) gains over the quadratic dependence
(\ref{excess_energy_canted}) of the canted phases at small doping. At
small doping then, the system is insulating and the additional
carriers are trapped in inhomogeneous lattice distortions. The
resulting phase is interesting in that it possesses some self-induced
disorder.  Random self-trapping of additional carriers has also been
reported in the adiabatic spinless Holstein model, with a
concentration of carriers close to one electron for two
sites,\cite{Pinaki} that is similar to the present case. By including
the double exchange, we have shown that such doping-induced
inhomogeneous states are in fact quite general and more stable than
the canted states because of the linear energy gain we have found at
small concentration. It has some similarities with the inhomogeneous
states with metallic droplets found in a simple model with
charge-ordering near half-filling.\cite{Kagan} We note that we have
considered here the simplest polaronic state with fully localised
electrons on the defect sites. It would be interesting to study
whether more complex defects, involving for instance distortions of
the magnetic structure on the neighbouring sites (magnetic polaron)
could arise near $x \sim 1/2$, as suggested for $x \sim 0$ (ref. [\onlinecite{LP-N}])
or $x \sim 1$ (ref. [\onlinecite{Meskine}]).

\begin{figure}[htbp]
\psfig{file=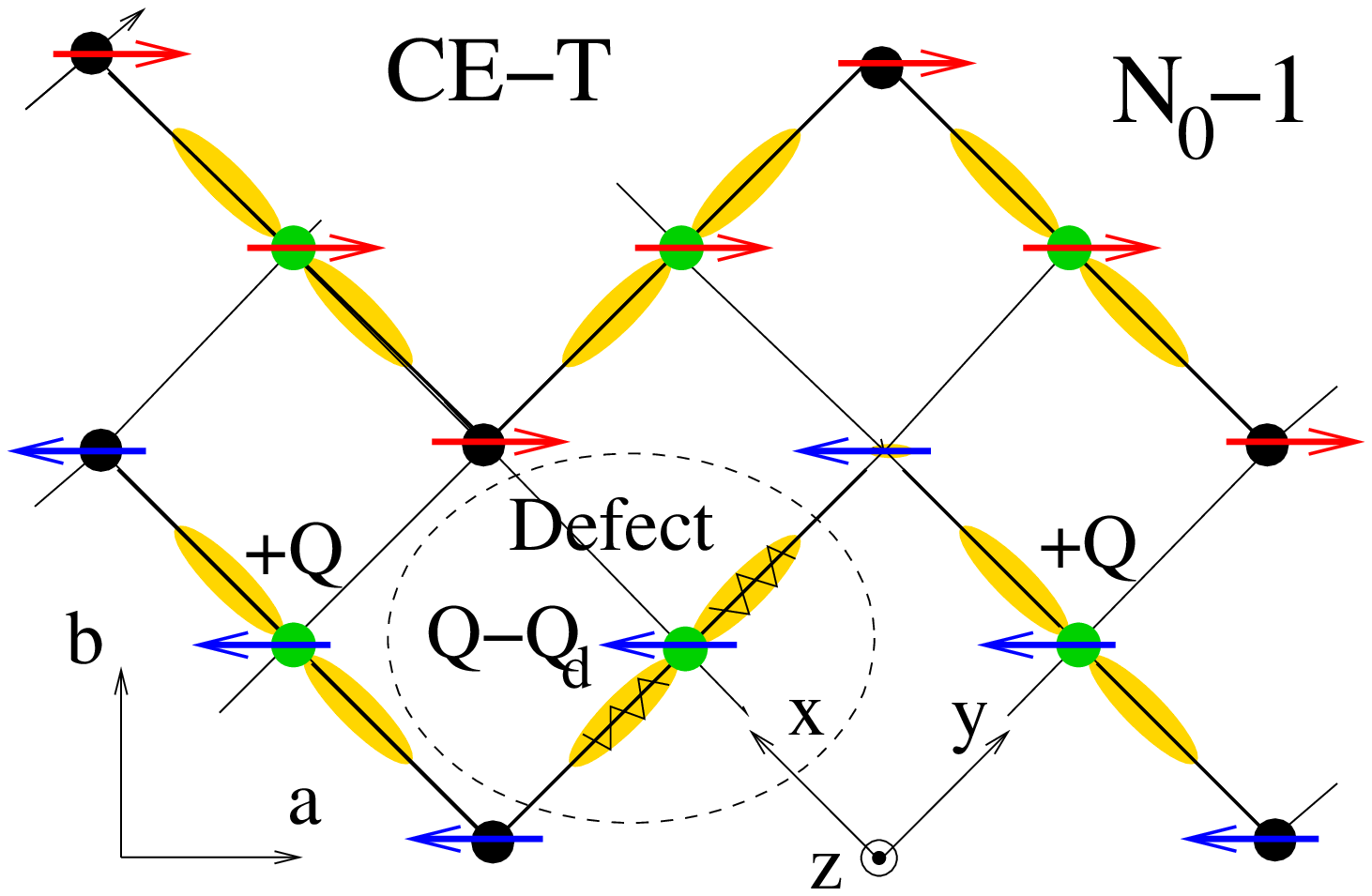,width=5cm,angle=-0}
\centerline{ 
\psfig{file=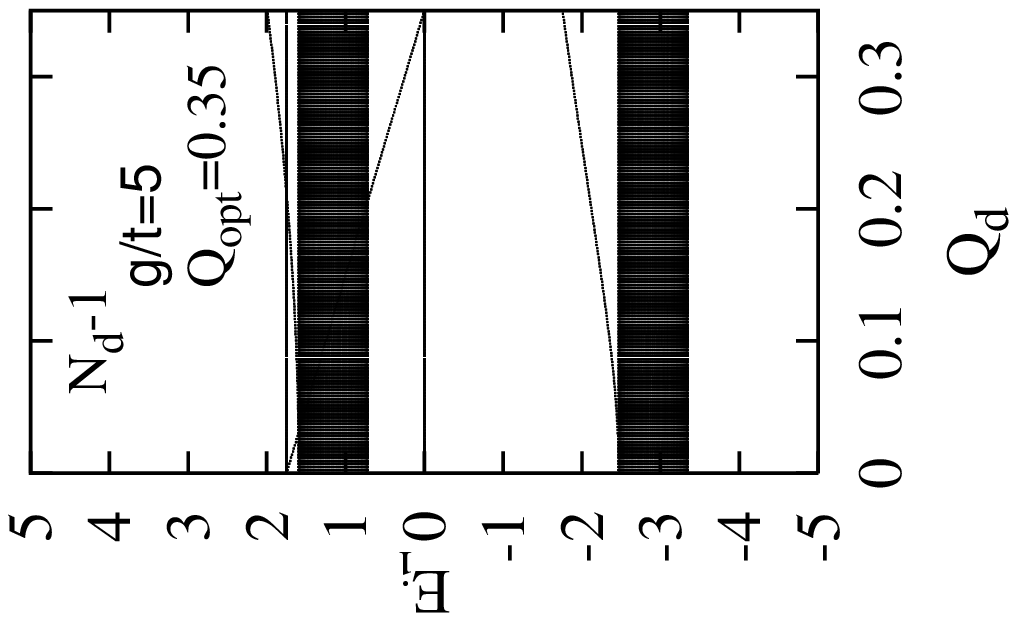,width=5cm,angle=-90}
\hspace{-0.7cm}
\psfig{file=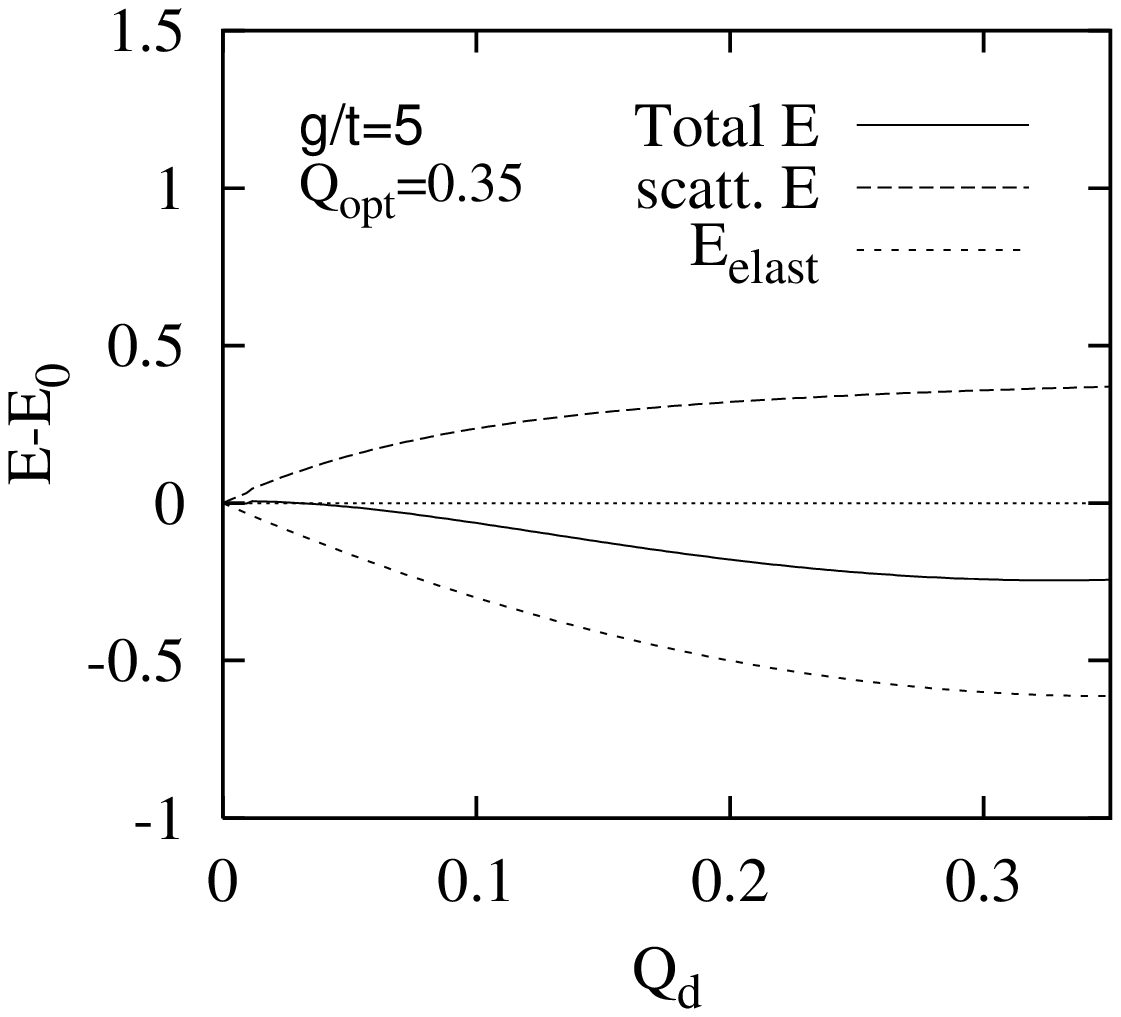,width=5cm,angle=-90}}
\caption{(color online). Definition of the defect on a single bridge
site with distortion decreased to $Q-Q_d$ ($Q_d=0$ corresponds to a
fully distorted site as in the original structure).  Energy levels
vs. $Q_d$ (bottom left) and energies for $N_0-1$ electrons (bottom
right).  $g/t=5$.}
\label{levelsdefecthole-f}
\end{figure}

\textit{Trapping of Added Holes:} Similarly, it is also favourable to
trap added holes in lattice distortions. We consider the analogue of
the previous problem with one-less distortion and $N_0-1$
electrons. The removal of the distortion on one site creates again a
defect. The energy levels and the excess energy calculated numerically
by filling the energy levels with $N_0-1$ electrons,

\begin{eqnarray}
E-E_0 = E_{el}(N_0-1,Q,Q_d) - E_{el}(N_0-1,Q,0) \nonumber \\  +
\frac{1}{2}K(Q-Q_d)^2 - \frac{1}{2}K Q^2
\label{ebsCE}
\end{eqnarray}
are given in Fig. \ref{levelsdefecthole-f}. We find that it is
favourable to trap the additional hole onto the bound state that
appears within the gap just above the lowest band (Fig.
\ref{levelsdefecthole-f}, left) as soon as $g/t \gtrsim 4$. The
energy gain, $\tilde{E}^h_{JT}$ is smaller than $E_{JT}$
and is $0.31t \sim 0.35E_{JT}$ for $g/t=5$ (Fig.
\ref{levelsdefecthole-f}). For the whole system the energy gain is
$- \tilde{E}_{JT}^h \delta c$. It is then favourable to trap the
holes and the system is insulating.

\subsubsection{Competition between the two phases and Phase Diagram}
\label{competionbetweenthetwo}

\begin{figure}[htbp]
\centerline{ \psfig{file=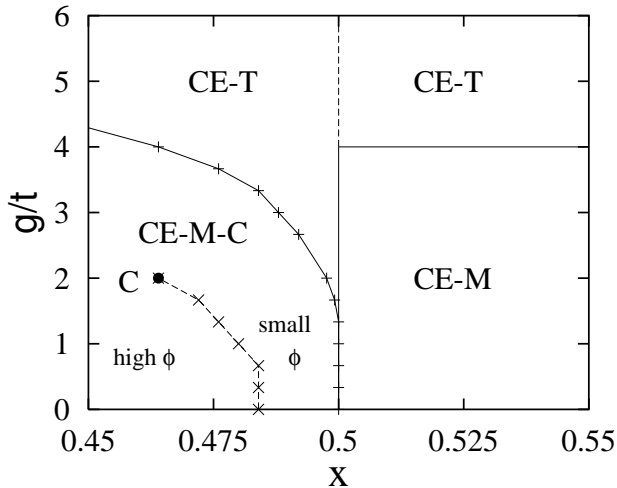,width=7cm,angle=-90}}
\caption{Phase diagram as a function of the model parameter $g/t$ for
hole doping concentration $x$ close to $1/2$
($J_{AF}S^2/t=0.15$). CE-M-C denotes a metallic and canted CE
phase. CE-T refers to the CE phase with added carriers
self-trapped in JT defects.\cite{notecurve} CE-M is the CE
phase with added holes that is metallic (no canting nor trapping of
the holes).  C is a critical point ending a first-order transition
line between two CE-M-C states with differing canting angles that
arise for small $g/t$.} \label{Phase diagram-f}
\end{figure}

By comparing the energies of the canted state and the defective state,
(for instance, as shown for $g/t =5$ in Fig.\ref{energiesdoping-f}) we
can arrive at a phase diagram in $g/t - x$ plane near half doping. As
discussed above, when electrons are added to the CE phase, they are
trapped on corner sites with newly generated Jahn-Teller distortions
if their concentration is within $[ 1/2~-~\delta c_{crit},~1/2]$ (Fig.
\ref{Phase diagram-f}). These JT distortions are oriented in the
$z$-direction, so as to favour the occupancy of the $(3z^2-r^2)$
orbital. The magnetic structure remains of CE type. For $x= 1/2 -
\delta c_{crit}$ there is a transition to a metallic state with canted
spins.  In the present approach there is a finite canting angle at the
transition, so that the transition is first-order.  When the
concentration increases further the canting angle becomes larger and
larger. At small $g/t$ we find that there is a first-order transition
line between a canted state with small canting angle and a highly
canted state that ends by a critical point (Fig. \ref{Phase
diagram-f}). The system becomes eventually fully ferromagnetic.

On the other hand, when holes are added to the CE phase, the holes get
trapped on bridge sites with lattice distortions removed when $g/t
\gtrsim 4$ (and the system remains insulating, CE-T) or move freely in
the lowest band for $g/t \lesssim 4$ (and the system is metallic,
CE-M) [Fig. \ref{Phase diagram-f}]. There is no competition with
canted states in this case because, as discussed in
\ref{doping-ss-canted}, the density of states near the top of the
valence band is not large enough to provide sizeable energy gains.

We emphasise that this phase diagram is based on an instability
calculation, and can be expected to be accurate only when $x$ is
sufficiently close to $x=1/2$. Away from it, the defects start to
interact and other phases may appear.\cite{incom-CO,Brey,Efremov}

We note that in all our discussions so far we have neglected the
disorder that is present in the real systems. In fact, even at
half-doping the arrangement of the A site ions (La or Ca) is
disordered. This disorder causes the localisation of the one electron
states (near the band-edges in 3-d) leading to increased stabilisation
of the insulating properties near half-doping, and may also have
consequences on the local magnetic structure as emphasised by the idea
of ferrons.\cite{LP-N}

\subsection{Instability of the CE Phase in a Magnetic-Field}
\label{mag-f-ss}

Finally, in this subsection we discuss the instabilities of the CE
phase in a magnetic-field due to the combined effects of canting of
spins towards the direction of the field by a canting angle $\phi$ (as
pictured in Fig.  \ref{latticePertu-f}) and modification of the JT
distortions.

Since a band opens out of the zero energy states upon canting (dashed
lines in Fig. \ref{bandstructure-f}), the system will "melt" into a
metallic phase when the corresponding charge gap closes. Since this
can happen for a canting angle less than $\pi/2$, the transition will
typically not be to the fully polarised ferromagnetic state.

\begin{figure}[htbp]
\centerline{ \psfig{file=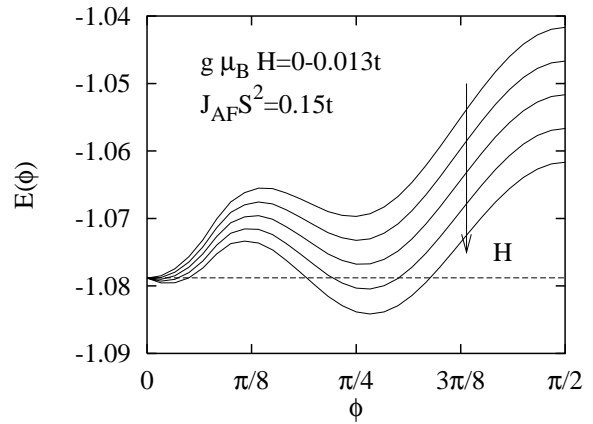,width=6cm,angle=-90}}
\caption{First-order transition of the CE phase to a canted state
when an external magnetic-field is applied to the system at $g=0$.
The threshold field that induces the first-order transition is
given by $\mathrm{g}\mu_B  H_c=0.010t$. The system is metallic
beyond $H_c$.} \label{cantingH-f}
\end{figure}

In Fig. \ref{cantingH-f}, we show the energy versus the canting
angle for different fields at $g/t=0$. There is a small optimal
canting angle that minimises the energy for small fields, but the
system continues to be insulating. When the field increases
further there is a first-order transition to a state with a finite
(and large) canting angle, which is metallic. This is reminiscent
of the first-order transition line between two canted states when
additional electrons are added (section \ref{doping-ss} and Fig.
\ref{Phase diagram-f}). We note that the first-order transition
field is very small, $\mathrm{g} \mu_B H_c=0.010t$, not only
because, with $J_{AF}S^2=0.15t$ the system is close to the phase
boundary, but also because the canted phases reduce the critical
field considerably. Previous works that have assumed that the
insulator-metal transitions involve the fully ferromagnetic
state\cite{Pandit,Fratini,Dagotto2} would predict a transition
field $\mathrm{g} \mu_B H_c=0.14t$ for the same $J_{AF}$ we have
used above. We discuss in detail next as to how these features
change with respect to turning on $g$ (and keeping the same
$J_{AF}$).

\subsubsection{$g/t \lesssim 5$}

\begin{figure}[htbp]
\centerline{ \psfig{file=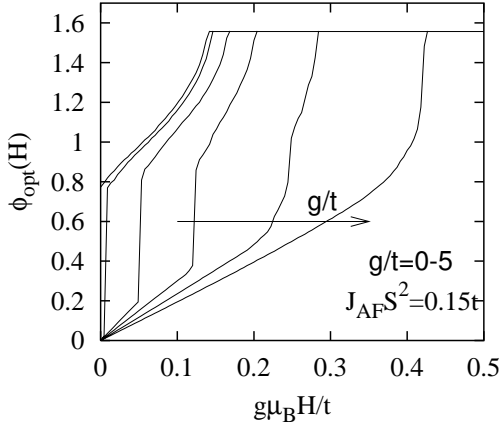,width=6cm,angle=-90}}
\caption{Optimal canting angle vs. external magnetic-field for
different values of $g/t$ (from 0 up to 5) {\it keeping the same
distortions} as at $H=0$ (CE phase) for all magnetic-fields. There
are first-order transitions between canted states. As we will
show, there are other transitions, involving modified JT
distortions, that preempt those shown in this figure (see below,
Fig. \ref{magnetizationgap-f}).  } \label{magnetization-f}
\end{figure}

To start with we take the CE phase with the optimal lattice
distortions found in section \ref{pd-s} for different values of
$g/t$. First, we {\it freeze the distortions} for all
magnetic-fields and find the optimal canting angle that minimises
the energy as a function of the magnetic-field, for non-zero
values of $g/t$ (Fig. \ref{magnetization-f}). The jumps in the
canting angle correspond to first-order transitions between states
with small and large canting angles. When $g/t$ is increased, the
transition fields and the fields at which the fully ferromagnetic
state is reached shift rather quickly to larger values.  The
transition field becomes larger simply because, with increasing
$g/t$ (and for the same $J_{AF}$), the system moves further and
further away from the phase boundary with the ferromagnetic phase
(see the phase boundary in Fig.  \ref{pd-f}).

\begin{figure}[htbp]
\centerline{ \psfig{file=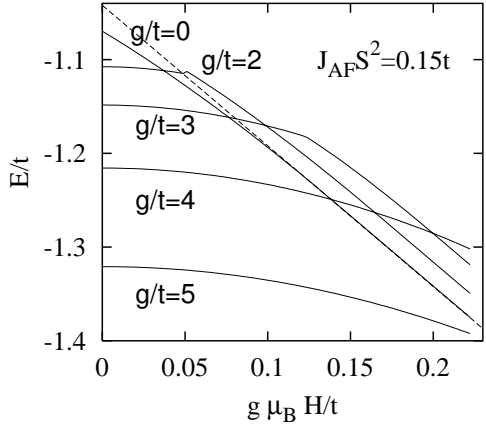,width=6cm,angle=-90}}
\caption{Crossing of the energies of the \textit{distorted} canted
CE phases (solid lines, with $g/t=2,3,4,5$) with that of the
\textit{undistorted} highly-canted phase (solid line corresponding
to $g/t = 0$), or the fully ferromagnetic phase (dashed line) to
which it merges for large fields. The cusps visible in the
$g/t=2,3$ curves at higher fields correspond to the first-order
transitions described in Fig. \ref{magnetization-f}. But these are
preempted by the transitions to the undistorted phase.}
\label{eh-f}
\end{figure}

\begin{figure}[htbp]
\psfig{file=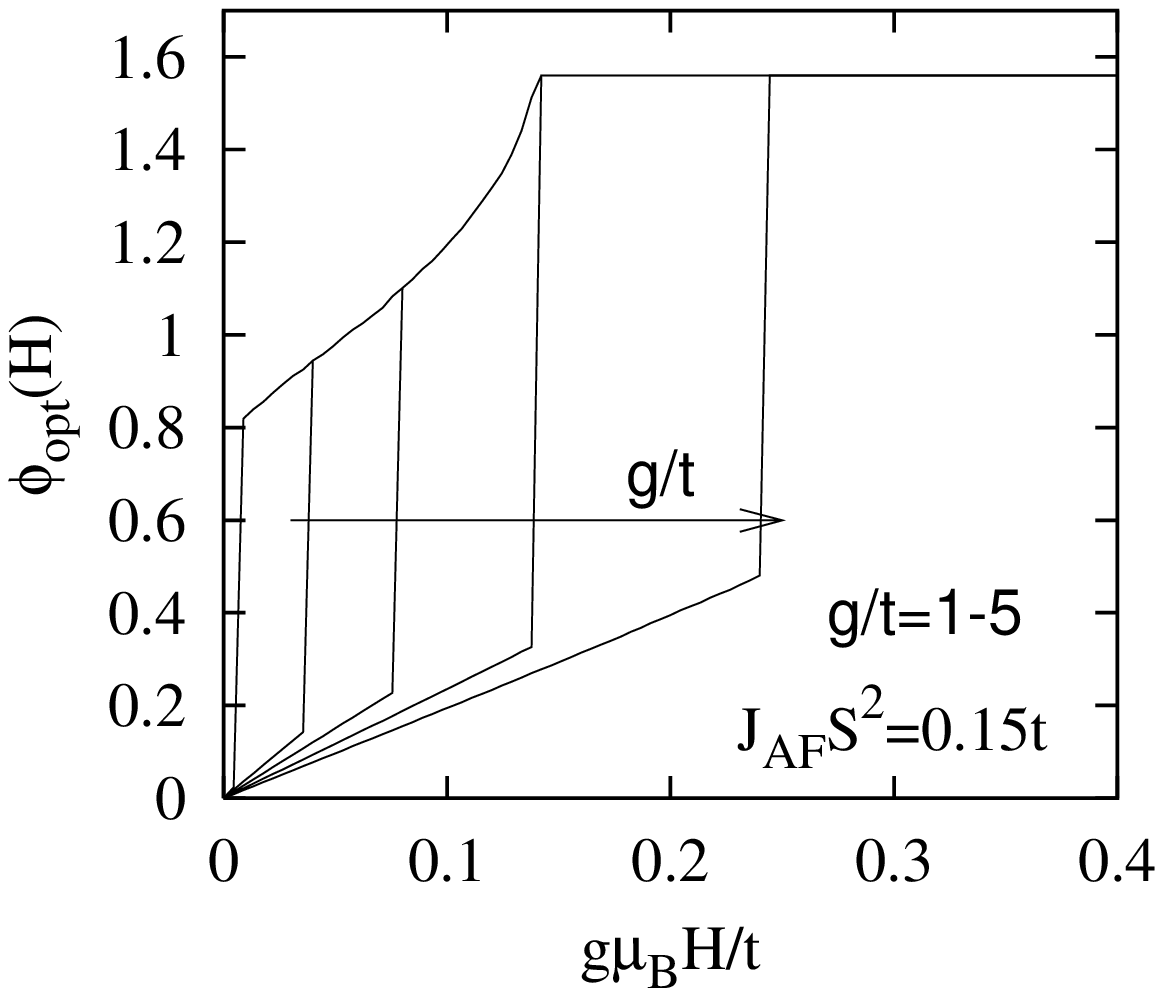,width=6cm,angle=-90} \\
\psfig{file=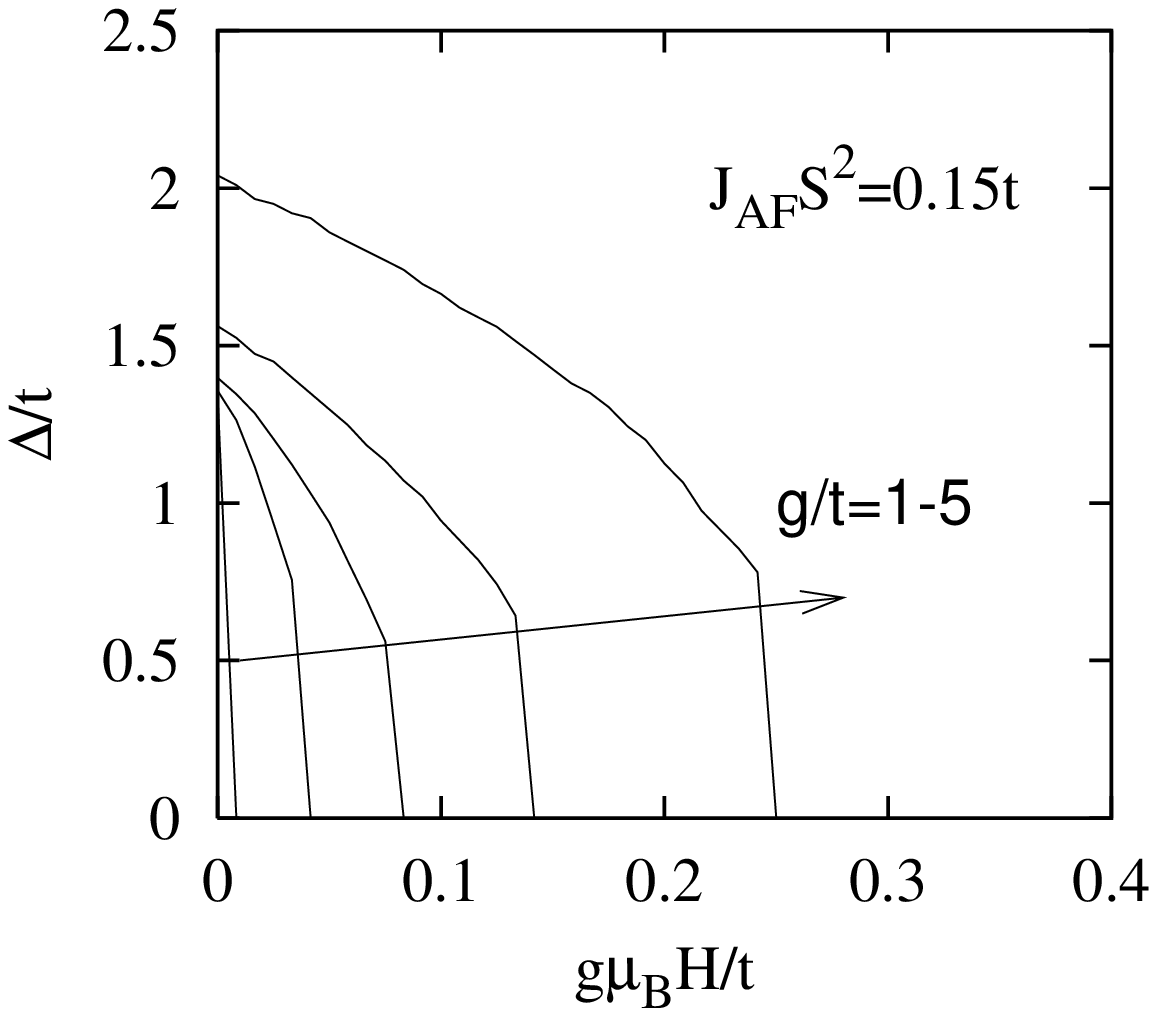,width=6cm,angle=-90}
\caption{Magnetisation or canting angle (top) and charge gap
(bottom) vs. external magnetic-field for different values of
$g/t=1-5$ ($J_{AF}S^2/t=0.15$). There are clear first-order
transitions to an  \textit{undistorted} canted metallic or fully
ferromagnetic metallic phase. } \label{magnetizationgap-f}
\end{figure}

We next show that the threshold fields for the "melting" of the CE
phase are reduced by taking into account the effect of the
magnetic-field on the lattice distortions themselves. One should
expect this, since in section \ref{pd-s} it was shown that the
ferromagnetic phases remains undistorted up to $g/t=5$. It is made
explicit in Fig. \ref{eh-f}, where the energy of the
\textit{undistorted} highly-canted state (solid line), which merge
into the fully ferromagnetic state (dashed line) at higher fields,
crosses that of the \textit{distorted} canted CE states. These
transitions occur at smaller transition fields (see Fig.  \ref{eh-f})
compared with the transition fields we considered in the previous
paragraph (the latter correspond to the cusps, visible in the curves
$g/t=2,3$ in Fig. \ref{eh-f}) .  The new magnetisation curves are
shown in Fig. \ref{magnetizationgap-f}, top. We then calculate the
band structure as a function of the magnetic field and extract the gap
(Fig. \ref{magnetizationgap-f}, bottom). The jumps in the
magnetisation and in the gaps turn out to be simultaneous, thus
indicating that the transitions correspond to insulator-metal
transitions. The transition fields $\mathrm{g}\mu_B H_c$ vary in the
range $0.01t-0.2t$ for $J_{AF}S^2=0.15t$. The smaller the $g/t$ the
smaller the transition field at fixed $J_{AF}$.

The first-order transitions discussed above were obtained by
considering the crossing of two solutions, namely the
\textit{distorted} canted CE phase (with the distortions frozen at
their $H=0$ values) and the \textit{undistorted} FM phase. It is
possible, in principle, that intermediate phases with intermediate
distortions appear. To rule out this possibility, we have
performed the full optimisation in presence of the external
magnetic-field for $g/t=4,5$ and found that the decrease in the
distortions is less than $6 \%$ up to the transition to the
undistorted ferromagnetic phase. This validates our approximation
of using frozen distortions in the canted CE phase up to the
first-order transition.

We can roughly understand some of the above results in terms of an
expansion of the physical quantities in the canting angle,
before the first-order transition takes place. For simplicity we
neglect the distortions on the corner sites of the CE state, so
that the zero-energy states are degenerate to start with. The
degeneracy is then lifted by the canting and the reduced charge
gap is given at first order in the canting angle by:
\begin{equation}
\Delta(\phi)=\Delta  - \beta_e t \phi
\label{gapphi}
\end{equation}
where $\Delta$ is the gap of the uncanted CE phase, given by equation
(\ref{gap}), and $\beta_e=8/3$ a numerical coefficient (from
eq. \ref{dispersionzeroband}). The canting angle itself is determined
by minimising the magnetic energy per spin due to canting, given for
small $\phi$ by:
\begin{equation}
E =\kappa  \phi^2  - \mathrm{g} \mu_B S H \phi
\end{equation}
where $\kappa \equiv 4J_{AF}S^2-\kappa_e$ is the spin stiffness we
have referred to before (section \ref{doping-ss}). The energy is
minimal for $\phi_{opt}(H) = \mathrm{g} \mu_B S H/2 \kappa$, which
describes the linear regime of the curves in Fig.
\ref{magnetizationgap-f}, before the first-order transitions to
the highly canted states take place. The corresponding charge gap,
given by
\begin{equation}
\Delta_{min} \equiv \Delta(\phi_{opt}) = \Delta - \left(\frac{\beta_e
t}{2\kappa}\right) \mathrm{g}\mu_B SH ,
\label{gapDH}
\end{equation}
also decreases linearly with field, but is not a very good
approximation to the results of Fig. \ref{magnetizationgap-f},
bottom. The difference comes from the small distortions on the corner
sites which we neglected in deriving (\ref{gapphi}) by assuming
degenerate zero-energy states. With small distortions on the corner
sites, the zero-energy states are no longer degenerate, and that would
modify the expression for the gap (\ref{gapphi}).

In any case, one can give an \textit{upper bound} for the critical
magnetic-field by extrapolating the linear expression (\ref{gapDH})
to zero. This is an upper bound because the actual transition occurs
before the gap get completely closed as is clear from
Fig. \ref{magnetizationgap-f}. We have

\begin{equation}
\mathrm{g} \mu_B S H_c =\frac{2 \kappa \Delta}{\beta_e t}=(8J_{AF}S^2 - 2 \kappa_e) \frac{\Delta}{\beta_e t}
\label{critical}
\end{equation}
It is clear from this expression that in order to close a gap
$\Delta$ of order $E_{JT}$ or $t$, we do not need a magnetic-field
of order $\Delta$ thanks to the reducing factor $\kappa/t$.
$\kappa/t$ is small, first because the spins are easy to polarise
on energy scales that have nothing to do with the charge scales,
as in the standard field-induced insulator-metal transition of a
spin-density wave.  It is interesting to note that the large field
$\mathrm{g}\mu_B S H_{c0} \equiv 8J_{AF}S^2$ is the critical field
to align antiferromagnetic spins coupled only by $J_{AF}$. The
real transition fields are substantially reduced compared to this
by the double exchange included in $\kappa_e$ and the factor
$\Delta/\beta_e t$ which describes how fast the gap closes with the
canting angle. For example, the strength of the transitions fields
of Fig. \ref{magnetizationgap-f} are in the range $\mathrm{g}\mu_B
H_c \sim 0.1t - 0.25t$. With $t \sim 0.2 $ eV, we have
$\mathrm{g}\mu_B H_c \sim 20 - 50 $ meV, or $H_c \sim 140 - 350$
Tesla, which are still much too large. We discuss this discrepancy
below. We now consider the other situation with $g/t \gtrsim 5.0$.

\subsubsection{Transition to an Inhomogeneous State}

For $g/t>6.8$ we have seen that even the ferromagnetic phases are
insulating. Hence it is to be expected that no insulator-metal
transition can take place in this regime. Instead, the CE phase
makes a transition to the ferromagnetic insulating phase
for sufficiently large fields, as is clear from the large $g/t$
limit phase diagram discussed in section \ref{wannier-ss}. For
$g/t<6.8$, however, we have shown earlier that it is favourable to
create defects and mobile electrons out of the FI-CO phase
(section \ref{InstabilityFerro-ss}). We expect therefore that
similar kinds of phases will be favoured by a magnetic-field as
well. We explore this issue further below, taking $g/t=6$ as an
example.

\begin{figure}[htbp]
\psfig{file=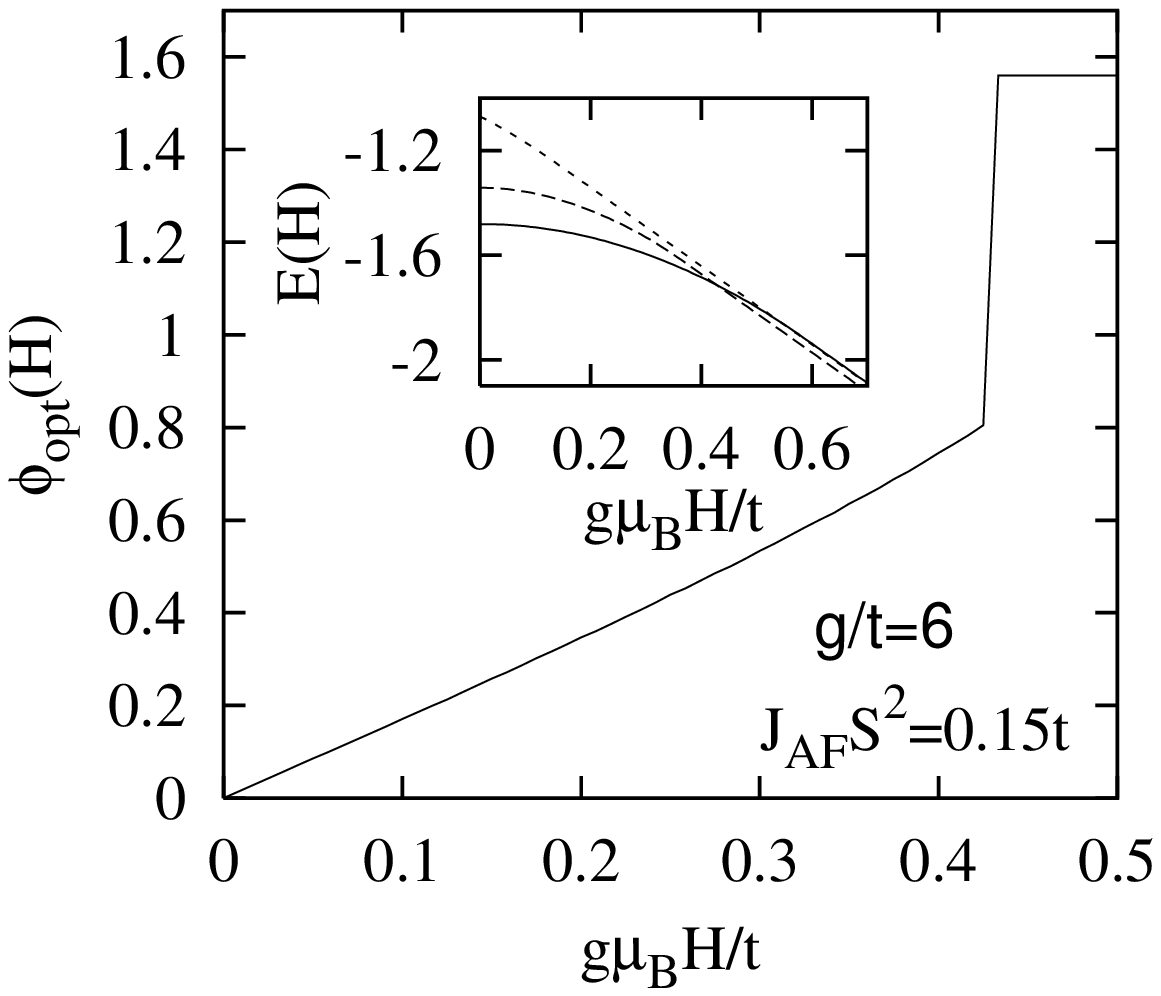,width=6cm,angle=-90} \\
\psfig{file=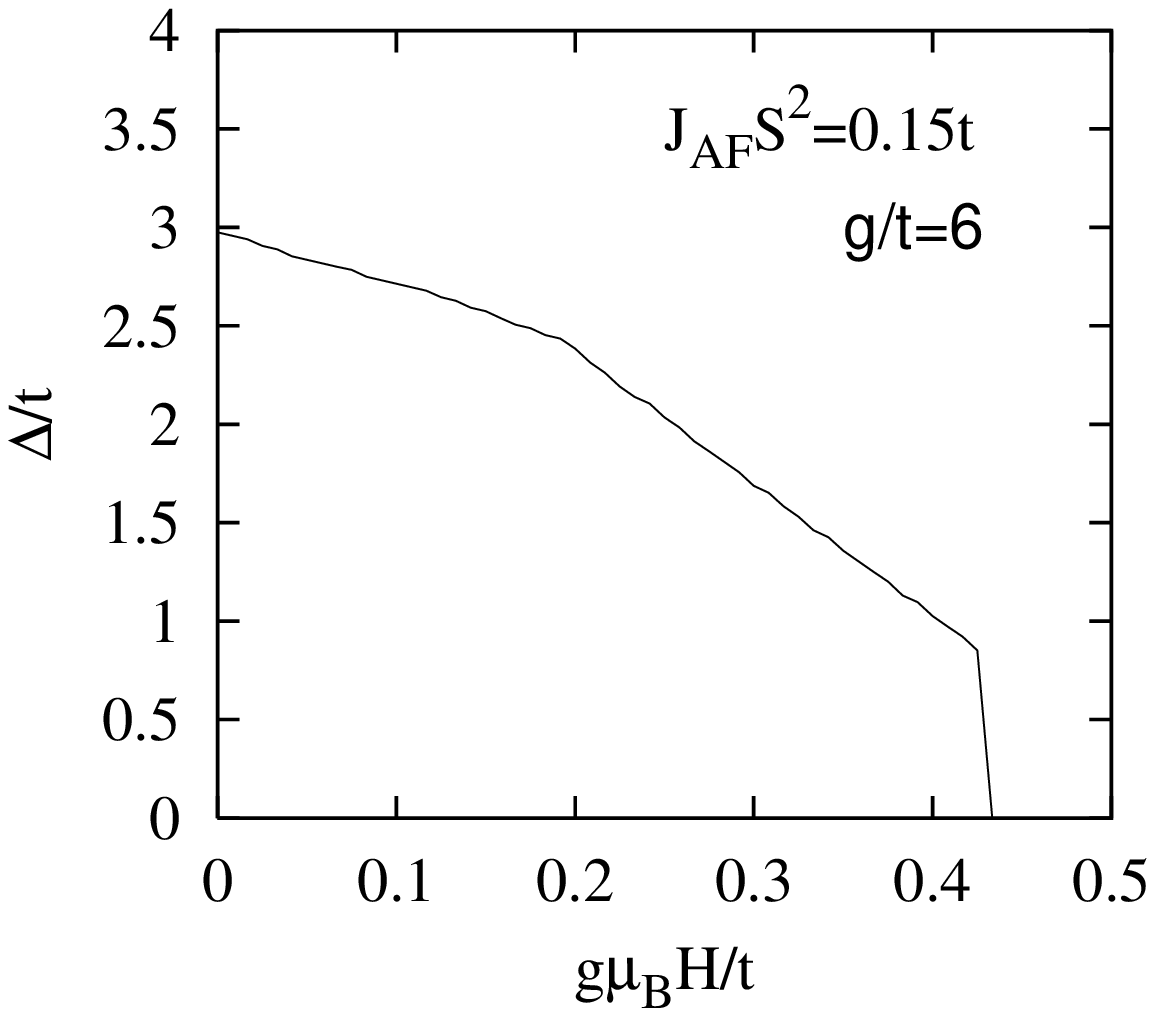,width=6cm,angle=-90}
\caption{Magnetisation or canting angle (top) vs. external
magnetic-field for $g/t=6$ ($J_{AF}S^2/t=0.15$). In the inset the
energies of the undistorted canted phase (dotted line), the canted
phase with distortions so as to favour $(x^2-y^2)$ orbital
ordering (dashed line) and the CE canted phase (solid line)  are
given. The latter two cross for $\mathrm{g}\mu_B  H /t $ around
0.43, where the first-order transition takes place. Above this
field the phase is fully ferromagnetic. Bottom: The resulting
Charge Gap is shown as a function of $\mathrm{g}\mu_B  H /t $}
\label{magnetizationgapg6-f}
\end{figure}

We first compare the energy of the canted distorted CE phase [with
the usual $(3x^2-r^2)/(3y^2-r^2)$ orbital ordering] with that of a
canted phase with alternate sites distorted in such a way as to
favour the $(x^2-y^2)$ orbital ordering. The latter is indeed a
lower energy state for large $g/t$ when the spins are fully
aligned, compared with the undistorted phase. We see in the inset
of Fig. \ref{magnetizationgapg6-f} that the two energies cross for
$\mathrm{g}\mu_B H /t \sim 0.43$. As shown in Fig.
\ref{magnetizationgapg6-f}, at this field the canting angle that
minimises the energy jumps from $\sim 0.8$ to $\pi/2$
corresponding to a transition to the fully ferromagnetic state
(which is metallic at $g/t=6.0$, see Fig. \ref{chargegap-f})  with
$(x^2-y^2)$ orbital ordering.

\begin{figure}[htbp]
\centerline{ \psfig{file=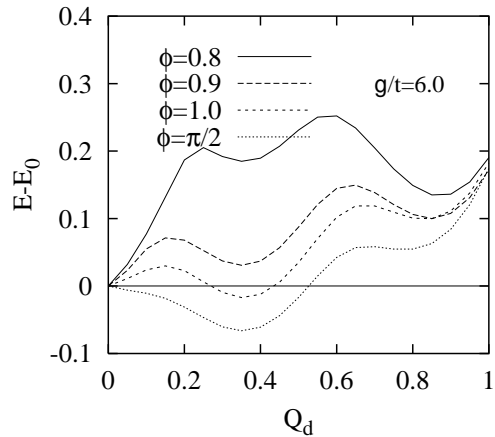,width=6cm,angle=-90}}
\caption{Energy for creating a single-site defect in the JT
distortions with amplitude $Q_d$ for various canting angles in the FI
phase. Such an excitation becomes soft at $\phi=0.95$ and the defects
proliferate.  The transition corresponds to an instability towards a
metallic phase with defects.} \label{energyinhomogeneous-f}
\end{figure}

We now consider the possibility of creating a defect in the
lattice distortion pattern, similarly to what we discussed in
section \ref{InstabilityFerro-ss}, which might lead to a lower
energy according to what we found there.  We start with the
$(x^2-y^2)$ phase with canting angle $\phi$, and reduce the
distortion of one of the distorted sites to $Q-Q_d$ instead of
$Q$, so that when $Q_d=Q$, the distortion is completely removed.
In Fig. \ref{energyinhomogeneous-f}, we show the total energy
$E-E_0$ (where $E_0$ is the energy of the homogeneous phase,
corresponding to $Q_d=0$) as function of $Q_d$ for different
canting angles near the transition. Clearly, there is a minimum at
$Q_d=Q$ ($Q=0.39$ at $g/t=6.0$) that corresponds to an excitation
where we remove a distortion and create a particle-hole
excitation. This minimum becomes soft at $\phi \sim 0.9$, thus
signalling the onset of a transition to a phase where such defects
are energetically favourable. Note that this phase can not
correspond to the undistorted phase (i.e., to removing the
distortions on all the sites) because the undistorted canted phase
is higher in energy as is clear from Fig.
\ref{magnetizationgapg6-f} (short-dashed lines).

To sum up the discussion above, when a magnetic-field is switched on,
the core spins cant towards the direction of the field. A band opens
out of the zero-energy states of the corner sites of the CE phase and
the gap gets reduced. At a threshold value of the magnetic-field,
there is an instability toward transferring electrons from the
localised states (or lower energy band) to the mobile states. For
field values close to this threshold, further energy gain is possible
when some Jahn-Teller distortions are removed and defects are
created. When the density increases, the missing distortions start to
play a more important role and the one-defect approach we have
developed here breaks down.  This scenario for the magnetic-field
induced metallic phase is very close to the description of the
metallic phase in the ($x<0.5$) regime as discussed in
Ref. [\onlinecite{Ram}].

Clearly, for all the values of $g/t$ we have considered, there are
rich transitions from the CE phase to the ferromagnetic phases as a
function of the magnetic-field. The lattice distortions play a
crucial role in converting a transition that would be naturally
second-order because of the progressive canting of the spins into
a first-order transition.

The strength of the magnetic-field at which the transition occurs,
although much reduced compared to naive estimates as discussed
above, is still much too large compared to experiment, especially
if we consider large values of $g/t$.  However, the large
distortions observed experimentally are consistent with relatively
large values of $g/t$.  How does one reconcile these two? As we
have seen, a crucial ingredient that determines the strength of
the transition field is the charge gap.  We can reasonably argue
that the gap is overestimated in the present approach. On the
experimental side, the gap is not of the order of $t$ but much
smaller, typically 5 times smaller for
Nd$_{1/2}$Ca$_{1/2}$MnO$_3$.\cite{Remark} This is also clear from
the temperature at which the charge-order transition takes place.
If we use the estimates of the gap from experiments in the
expression (\ref{critical}), the transition field comes out to be
much smaller, of order 30 T, much closer to the transition fields
of that compound.

It is clearly important, therefore, to improve the theory
presented above to generate a more accurate estimate for the
charge gap in the CE phase. For this we need to include at least
three sources of correction to the gap estimate above. Namely the
finiteness of the Hund's coupling $J_H$, the cooperative nature of
the JT distortions, and small second neighbour hoppings. When
$J_H$ is finite, the electrons can hop even between sites which
have anti-aligned core spins. This opens a band out of the zero
energy states and reduces the gap. Second, because distortions of
neighbouring sites are coupled, a distortion on a site imposes a
distortion on the neighbouring site that lowers the energy level,
thereby reducing the gap. Finally, second neighbour hopping allows
the corner site electrons to become mobile (even in the case of
infinite $J_H$), and overall increases the bandwidths of all the
bands, hence reducing the charge gap (see, for example,  ref.
[\onlinecite{Popovic}]). We believe that, while the theory
presented above clarifies how the smallness of the transition
field arises, a more elaborate theory including these three
effects which reduce the gap is required to obtain a precise
estimate of the transition field.

\section{Concluding Remarks}
\label{Conclusion-s}

In conclusion, we have first confirmed various periodic phases of the
phase diagram of half-doped manganites (Fig. \ref{pd-f}), by
optimising the lattice and magnetic energies in the thermodynamic
limit. We have thus provided explicit calculations for the Jahn-Teller
distortions, charge and orbital order-parameters in the various stable
phases by exploring the phase space in an unbiased way, albeit limited
by the 8-site unit-cell.

It is interesting to discuss, in the context of the phase diagram of
Fig.  \ref{pd-f}, the strengths of the couplings of real manganites
materials. Obviously, the absence of 3d ferromagnetic phases at
half-doping suggests that $J_{AF}$ is substantial in these materials
($J_{AF}S^2/t \gtrsim 0.05$).  The observation of the A-type phase in
Pr$_{1/2}$Sr$_{1/2}$MnO$_3$ with distortions that favour the
$(x^2-y^2)$ orbital,\cite{Kawano0,Kawano} is only compatible with $g/t
\gtrsim 5.0$ (Fig. \ref{pd-f}), a large value that is generally
corroborated by large distortions.\cite{Radaelli} In addition, the
fact that many half-doped manganites show a (CE) insulator to (ferro)
metal transition as a function of magnetic-field confirms that $g/t$
can not be much larger (since for $g/t \gtrsim 6.3$, the ferromagnetic
phase is also insulating). Adopting such values $g/t \sim 5$ (or
$E_{JT} \sim t$), the charge disproportionation in the CE phase
is $\delta \sim 0.2$ (Fig. \ref{chargedisproportionation-f}), which is
much smaller than Goodenough's ionic picture value of 0.5. The
inclusion of cooperative JT effects is likely to reduce the charge
contrast. In addition, the specific distribution of the charge
contrast amongst the Mn and O orbitals depends upon band-structure
details which we have not considered in this paper.\cite{low-en-p-H}

Secondly, and more importantly, we have studied the instabilities of
these phases with respect to canting of spins {\it and} single site
defects in their JT distortion pattern, caused by doping away from
$x=0.5$ or by the application of a magnetic-field. The consideration
of canted CE phases allowed us to study how the magnetisation changes
with applied field. We have found that the distortions do not change
much in the linear regime of the magnetisation up to a threshold field
at which there is an abrupt change. This seems to be consistent with
recent experiments.\cite{Tyson,Nojiriprivate} A more detailed
comparison would be interesting in order to extract the strength of
the JT coupling. Regarding the effect of the doping, we have found
that when electrons are added (with respect to half-doping, i.e., for
$x < 0.5$) the transition from an insulating CE phase with
self-trapped carriers to the ferromagnetic metallic phase of the
colossal magneto-resistance materials proceeds via a first-order
transition to canted states. In contrast, added holes (corresponding
to $x > 0.5$), do not favour canted states because of a lack of
density of states near the top of the valence band.  The holes prefer
to be self-trapped, at least above a threshold $g/t$. This, we
believe, is the underlying cause of why the most manganites tend to
remain insulating and favour incommensurate charge
order\cite{incom-CO,Brey} for $x > 0.5$, but quickly become metallic
for $x < 0.5$. This striking particle-hole asymmetry has been recently
explored within a Ginzburg-Landau framework\cite{milward}; we believe
that our work clarifies the microscopic basis for this
asymmetry.\cite{brey-plb-sol}

We emphasise that the actual numbers we have obtained, such as the
threshold $g/t$ and magnetic-field values for the various transitions
may not correspond to experimental data because the model we have
studied neglects several important effects such as the cooperative
nature of the JT distortions, the finiteness of the Hund's coupling,
presence of second neighbour hopping, etc. As pointed out in the
previous section, for example, the size of the charge gap in the CE
phase, which in turn determines the threshold field for its transition
to the ferromagnetic metallic phase, will be reduced when these
effects are taken into account.

Finally, and most importantly, in the intermediate JT coupling
regime which we have argued is  the most appropriate for
manganites, we have shown that there is an instability of the
ferromagnetic phase to formation of defects in the JT distortion
pattern. We did this by calculating the energy cost for creating a
single site defect in the lattice distortion; i.e. reducing (and
eventually removing) the distortion at that site and promoting a
quasi-localised electron onto a mobile band. We have shown that
there are parameter regimes where this appears spontaneously. A
proliferation of such defects leads to a scenario for the
ferromagnetic metallic state that is completely consistent with,
and provides a new justification for, the effective two-fluid (one
light and extended, the other polaronic and localised) picture
proposed recently to explain the insulator-metal transition in the
colossal magneto-resistance materials.\cite{Ram}

On the basis of this identification we have suggested a new effective
Hamiltonian given by eq. (\ref{effectiveEmobile}) which goes beyond
that of Ref. [\onlinecite{Ram}] in that it allows for possible orbital
and charge ordering effects. We believe that a treatment of this
Hamiltonian using more sophisticated methods, such as the dynamical
mean field theory,\cite{dmft-rev} will eventually lead to a more
complete theory of the manganites, including orbital and charge
ordering effects. We hope to discuss such work elsewhere.

Nevertheless,
our work suggests that the ferromagnetic metallic phase obtained
at large magnetic-fields in half-doped manganites is similar to
the ferro-metallic phases found upon hole doping, i.e., for
$x<0.5$, except, perhaps for some remanent orbital and charge
order. It is an obvious and interesting question as to whether any
vestige of the orbital and charge order present in the CE phase
survives metallisation. If it does, the metallic phase would also
be rather anisotropic, with larger mobility along the z-axis of
the CE phase. But irrespective of this, it should have a large
fraction of sites which continue to be JT distorted accounting for
the majority of the $e_g$ electrons, which remain localised, and
only a small number of mobile carriers. Interestingly, this resembles the
phenomenological picture of ref. \onlinecite{Roy}. Our results
provide a microscopic justification for this picture and can be
further tested experimentally in a variety of other ways, such as
measurement of Drude weights in optical conductivity, EXAFS and
neutron diffraction experiments, for instance.

\begin{acknowledgments}
O.C. would like to thank G. Bouzerar, T. Chatterji, G. Jackeli,
D. Khomskii, Y. Motome, H. Nojiri and T. Ziman for stimulating
discussions.  O.C acknowledges financial supports from the I.L.L. and
from the Indo-French grant IFCPAR/2404.1.
\end{acknowledgments}

\appendix

\section{Wave-functions of the CE state with lattice distortions}

The unit-cell of the 1d zig-zag chains has four inequivalent sites and
there are two orbitals per site (see Fig. \ref{CE-phase-f}).  We list
below the eight states, which extends
Refs. [\onlinecite{Terakura,Khomskii,Jackeli}] to the presence of
lattice distortions on the bridge sites. The corresponding energies
are given in Fig. \ref{bandstructure-f} and in equations (\ref{bseq1}) -
(\ref{bseq2}).
\begin{eqnarray}
&(i)& \hspace{0.2cm} \Psi^{\pm}_{q_a,1} = \frac{A^{\pm}_{q_a1}}{\sqrt{2}}  \left[ | 1,3x^2-r^2
\rangle_{q_a} - e^{iq_a} |3,3y^2-r^2 \rangle_{q_a} \right]  \nonumber \\
&-& A^{\pm}_{q_a1} \frac{2\sqrt{2}t}{3 \epsilon_{q_a1}^{\pm}} \left[| 2,3x^2-r^2 \rangle_{q_a}
+  e^{iq_a}  | 2,3y^2-r^2 \rangle_{q_a} \right] \nonumber \\
&-& A^{\pm}_{q_a1} \frac{2\sqrt{2}t}{3 \epsilon_{q_a1}^{\pm}} e^{iq_a} \left[ | 4,3y^2-r^2
\rangle_{q_a} + e^{iq_a}  | 4,3x^2-r^2 \rangle_{q_a} \right]
\nonumber \\
&& \epsilon_{q_a1}^{\pm} = -E_{JT} \pm \sqrt{E_{JT}^2 + \tilde{t}^2 (2+\cos q_a)}   \nonumber \\
&& A^{\pm}_{q_a1} = \left[1+\frac{\tilde{t}^2 (2+\cos q_a)}{(\epsilon_{q_a1}^{\pm})^2} \right]^{-1/2} \nonumber
\end{eqnarray}

\begin{eqnarray}
&(ii)& \hspace{0.2cm} \Psi^{\pm}_{q_a,2} = \frac{A^{\pm}_{q_a2}}{\sqrt{2}}  \left[ | 1,3x^2-r^2 \rangle_{q_a}
+ e^{iq_a} |3,3y^2-r^2 \rangle_{q_a} \right]  \nonumber \\
&-& A^{\pm}_{q_a2} \frac{2\sqrt{2}t}{3 \epsilon_{q_a1}^{\pm}} \left[| 2,3x^2-r^2 \rangle_{q_a} -  e^{iq_a}
| 2,3y^2-r^2 \rangle_{q_a} \right] \nonumber \\
&-& A^{\pm}_{q_a2} \frac{2\sqrt{2}t}{3 \epsilon_{q_a1}^{\pm}} e^{iq_a} \left[-| 4,3y^2-r^2
\rangle_{q_a} + e^{iq_a}  | 4,3x^2-r^2 \rangle_{q_a} \right] \nonumber \\
&& \epsilon_{q_a2}^{\pm} = -E_{JT} \pm \sqrt{E_{JT}^2 + \tilde{t}^2 (2-\cos q_a)}   \nonumber \\
&& A^{\pm}_{q_a2} = \left[1+\frac{\tilde{t}^2 (2-\cos q_a)}{(\epsilon_{q_a2}^{\pm})^2} \right]^{-1/2} \nonumber
\end{eqnarray}

\begin{eqnarray}
&(iii)& \hspace{0.2cm} \Psi_{q_a, 3} =  \frac{A_{q_a 3}}{\sqrt{2}} \left[  | 2,y^2-z^2 \rangle_{q_a}
+ e^{-iq_a} | 2,x^2-z^2 \rangle_{q_a} \right] \nonumber \\
&& - (2 \rightarrow 4, q_a \rightarrow -q_a) \nonumber \\
&& \hspace{0.2cm} \Psi_{q_a, 4} =   \frac{A_{q_a 4}}{\sqrt{2}} \left[  | 2,y^2-z^2 \rangle_{q_a} - e^{-iq_a}
 | 2,x^2-z^2 \rangle_{q_a} \right]  \nonumber \\
&-& (2 \rightarrow 4, q_a \rightarrow -q_a) \nonumber \\
&& A_{q_a 3} = (2 - \cos q_a)^{-1/2}; \hspace{0.6cm} A_{q_a 4} = (2 + \cos q_a)^{-1/2} \nonumber \\
&& \epsilon_{3,4} = 0 \nonumber
\end{eqnarray}

\begin{eqnarray}
\hspace{-1cm} &(iv)& \hspace{0.2cm} \Psi_{q_a, 5} =  | 1,y^2-z^2 \rangle_{q_a} \nonumber \\
&&  \Psi_{q_a, 6} = | 3,x^2-z^2 \rangle_{q_a} \nonumber \\
&& \epsilon_{5,6} = 2E_{JT}
\nonumber
\end{eqnarray}
where the notations are $\tilde{t}=4t/3$, $E_{JT}=gQ/2$ and the states are defined by:

\begin{eqnarray}
|a,\alpha \rangle_{q_a} = \frac{1}{\sqrt{N}} \sum_i e^{iq_aR_i} |i,a,\alpha \rangle \nonumber
\end{eqnarray}
with $a=1,..,4$ the four inequivalent sites of the zig-zag chain and
$\alpha$ can be any of the orbital states defined in
eq. (\ref{orbitaldefinition}).  $q_a$ is the component of the
wave-vector along the chain direction. It takes values in the first
Brillouin zone $[-\pi/2,\pi/2]$. As for $g=0$, the degeneracies at
$\pm \pi/2$ come from the translation symmetry combined with a mirror
plane symmetry. Similarly, the properties $\Psi_{q_a,
1}^{\pm}=\Psi_{q_a+\pi, 2}^{\pm}$ and $\Psi_{q_a, 3}=\Psi_{q_a+\pi,
4}$ are consequences of the same symmetry.


\begin{thebibliography}{99}
\bibitem{noteKhomskii} e.g., in Ref. \onlinecite{Efremov}, it is shown that a similar
phase, although not identical as far as the magnetic structure is
concerned, has a lower energy when electrons are added to the CE
phase. The CE phase is, however, still preferred at half-doping.
\bibitem{LP-comm} While this projection for the
$J_H \rightarrow \infty$ limit obtained by approximating the
$t_{2g}$ spins as classical is widely used, an approach valid for
quantum $t_{2g}$ spins using a representation in which the total
spin $(\vec{s}_i + \vec{S}_i)$ is diagonal is discussed in the
review article by Loktev and Pogorelov [\onlinecite{LP-N}].
\bibitem{hopping} A compact way of writing the hopping matrix
elements is in the form of an outer product of orbital
pseudo-spinor wavefunctions: $$t_{ab ij}^{\alpha \beta} = (4t/3)
\zeta_{ab ij}^{\alpha} \zeta_{ab ij}^{\beta}$$ The pseudo-spinors
corresponding to hopping in the 3 cubic directions are,
$$ \zeta_{x} = (\sqrt{3}/2, -1/2); \zeta_{y} = (-\sqrt{3}/2, -1/2); \zeta_{z} = (0,1).$$
\bibitem{notedifference} Note that the difference between the critical
values of $J_{AF}S^2/t$ at $g=0$ with that of Refs.
[\onlinecite{Khomskii,Jackeli}] with $V=0$ comes from a different
definition of $t$ by a factor $4/3$ but the results perfectly agree
otherwise.
\bibitem{noteneglectedterms} In writing this, we have neglected terms
proportional to $ n_i n_j$ since, as we have seen, for the half doped
case and for large $g/t$ there is nearly complete charge
disproportionation, whence they are smaller by a factor of $t/E_{JT}$
compared to the term retained above even when $U$ is zero, and are
completely suppressed for large $U$.
\bibitem{remark} We note that it would be important to take into account
the cooperative nature of the distortions if one wants to give a
quantitative account of the energies. This would in particular
modify the distortions of the corner sites. Since we are using the
simpler framework of non-cooperative JT phonons, we therefore
neglect the small distortions of the corner sites also at this
stage.
\bibitem{notecurve} The transition line between the CE-M-C and the
CE-T phases vanishes at $x=1/2$, as seen in the figure, because the
conduction band in the CE phase is dispersionless in the nearest
neighbor model we have considered. With a more realistic model, the
transition would occur at a finite $g/t$ at $x=1/2$.
\bibitem{Remark} We can fit the resistivity-temperature curve to
get an estimation of the gap. We find a gap of the order of $500$
K from Ref. [\onlinecite{Millange}], which is much smaller than
$t$.
\end{thebibliography}
\end{document}